\documentclass[english]{article}
\usepackage[T1]{fontenc}
\usepackage[utf8]{inputenc}
\usepackage{geometry}
\usepackage{color}
\usepackage{babel}
\usepackage{float}
\usepackage{amsmath}
\usepackage{graphicx}
\usepackage{setspace}
\usepackage[unicode=true,pdfusetitle,
 bookmarks=true,bookmarksnumbered=false,bookmarksopen=false,
 breaklinks=false,pdfborder={0 0 1},backref=false,colorlinks=false]
 {hyperref}

\makeatletter

\newcommand{\lyxdot}{.}

\makeatother

\begin{document}
\begin{doublespace}
\begin{center}
\textbf{\textcolor{black}{\large{}Hong Kong - Shanghai Connect / Hong
Kong - Beijing Disconnect (?):}}{\large\par}
\par\end{center}

\begin{center}
\textbf{\textcolor{black}{\large{}Scaling the Great Wall of Chinese
Securities Trading Costs}}{\large\par}
\par\end{center}

\begin{center}
\textbf{Ravi Kashyap }
\par\end{center}

\begin{center}
\textbf{SolBridge International School of Business / City University
of Hong Kong }
\par\end{center}

\begin{center}
\begin{center}
\today
\par\end{center}
\par\end{center}

\begin{center}
Hong Kong; Shanghai; Connect; Trading Cost; Market Impact; Uncertainty
\par\end{center}

\begin{center}
Forecasting and Simulation
\par\end{center}
\end{doublespace}

\begin{center}
\textbf{\textcolor{blue}{\href{http://www.iijournals.com/doi/abs/10.3905/jot.2016.11.3.081}{Edited Version: Kashyap, R. (2016). Hong Kong–Shanghai Connect/Hong Kong–Beijing Disconnect? Scaling the Great Wall of Chinese Securities Trading Costs. The Journal of Trading, 11(3), 81-134. }}}\tableofcontents{}
\par\end{center}
\begin{doublespace}

\section{Abstract }
\end{doublespace}

\begin{doublespace}
We utilize a fundamentally different model of trading costs to look
at the effect of the opening of the Hong Kong Shanghai Connect that
links the stock exchanges in the two cities, arguably the biggest
event in international business and finance since Christopher Columbus
set sail for India. We design a novel methodology that compensates
for the lack of data on trading costs in China. We estimate trading
costs across similar positions on the dual listed set of securities
in Hong Kong and China, hoping to provide useful pieces of information
to help scale \textbf{\textit{“The Great Wall of Chinese Securities
Trading Costs”}}. We then compare actual and estimated trading costs
on a sample of real orders across the Hong Kong securities in the
dual listed pair to establish the accuracy of our measurements.

The primary question we seek to address is \textbf{\textit{“Which
market would be better to trade to gain exposure to the same (or similar)
set of securities or sectors?”}} We find that trading costs on Shanghai,
which might have been lower than Hong Kong, might have become higher
leading up to the Connect. What remains to be seen is whether this
increase in trading costs is a temporary equilibrium due to the frenzy
to gain exposure to Chinese securities or whether this phenomenon
will persist once the two markets start becoming more and more tightly
coupled. 

It would be interesting to see if this pioneering policy will lead
to securities exchanges across the globe linking up one another, creating
a trade anything, anywhere and anytime marketplace. Looking beyond
mere trading costs, such studies can be used to gather some evidence
on what effect the mode of governance and other aspects of life in
one country have on another country, once they start joining up their
financial markets.
\end{doublespace}
\begin{doublespace}

\section{Introduction}
\end{doublespace}

\begin{doublespace}
On November 17, 2014, amidst the backdrop of the protests in Hong
Kong regarding electoral reform, the plan to connect the stock markets
of Hong Kong and Shanghai proceeded after a slight delay over the
preceding weeks. The opening of the Hong Kong - Shanghai connect,
henceforth “ Connect”, opens a new era in the cross border flow of
capital into and out of China. While the proximate intention behind
this scheme could be to increase the trading of securities and bolster
the equity markets in China, the fundamental reasoning could be to
liberalize the financial system and spur economic growth, which has
fallen sharply from the double digit rates of the recent past (End-note
\ref{enu:A-Review-of}). Whether this is part of a bigger scheme to
financially join the two economies and aid greater political unification
is a matter to be studied over the next few decades. Also of interest
would be to see if this pioneering policy will lead to securities
exchanges across the globe linking up one another, creating a trade
anything, anywhere and anytime financial marketplace.
\end{doublespace}
\begin{doublespace}

\subsection{Stock Markets and Economic Growth }
\end{doublespace}

\begin{doublespace}
There is a vast body of literature regarding financial liberalization
and economic growth. Bekaert, Harvey and Lundblad (2005) show that
equity market liberalizations on average lead to an increase in annual
real economic growth rates. They point out that equity market liberalization
directly reduces financing constraints by making available more foreign
capital, and foreign investors may insist on better corporate governance,
which should promote financial development. Levine (2001) suggests
that stock markets influence growth through efficient capital allocation.
New information can lead to profitable trading and improved information
about firms improves resource allocation and hence economic growth. 

Levine and Zervos (1996, 1998a) find that stock market liquidity –
as measured both by the value of stock trading relative to the size
of the market and by the value of trading relative to the size of
the economy – is positively and significantly correlated with current
and future rates of economic growth, capital accumulation and productivity
growth. Levine and Zervos (1998b) looks at the stock market liquidity
following capital control liberalization in 15 emerging market economies
and find that stock markets tend to become larger, more liquid, more
volatile and more integrated following liberalization. Building on
the consensus that stock market liberalizations can reduce the cost
of equity capital, Henry (2000), finds that there can be a boom in
private investments following the liberalization. Beck and Levine
(2004) show that stock markets are important for economic growth independent
of the banking sector, with some evidence that the two could provide
different set of financial services and could be complimentary to
one other. Deeg and O`Sullivan (2009) chronicle the shift from predominantly
state controlled financial systems to multilateral agencies like the
International Monetary Fund. They emphasize the increasing significance
of regulatory regimes generated through the interactions of public
and private actors that extend across national boundaries. The discussion
turns towards the consequences of the current trends toward financialization
and the recovery that is currently underway after the 2008 collapse.
Epstein (2005) defines financialization, broadly, as the increasing
role of financial motives, financial markets, financial actors and
financial institutions in the operation of the domestic and international
economies; and goes on to a deeper discussion regarding the dimensions
of financialization, its implications for economic stability, growth,
income distribution, political power, policy formulation. Adding to
the importance of liquidity for asset pricing, Bekaert, Harvey and
Lundblad (2007) use measures of liquidity constructed using daily
returns and the length of the non-trading interval to show that liquidity
predicts future returns. While admitting that transaction data are
hard to obtain in emerging markets, they justify the usage of this
alternate measure for emerging markets. 

One aspect that stands out among all these studies is that none of
them explicitly consider the implicit trading costs or the uncertainty
associated with market prices, either in the measures of liquidity
or otherwise, while the actual transfer of capital happens through
the means of trading in the stock markets. In this study, we focus
heavily on the trading cost element when there is a major liberalization
event. We utilize a fundamentally different method of measuring trading
costs and apply it to study one of the, if not the, biggest event
in international business and finance since Christopher Columbus set
sail for India. If we find that trading costs are being influenced
heavily by market liberalizations, the lessons from such a study can
be manifold, change in transaction costs could be a signal of potential
building up of a bubble and a later bust. When not indicative of such
extreme situations, trading or transaction costs, could serve to highlight
whether the transfer of capital or investment is happening as expected
and whether foreign investors view on liquidity has been shaped positively,
to bring in additional sources of capital.
\end{doublespace}
\begin{doublespace}

\subsection{The Connect }
\end{doublespace}

\begin{doublespace}
The official announcement of the “Connect” program launch on April
10, 2014, sent financial intermediaries into an arms race, of sorts,
to be prepared to trade on the connect right from the first day of
the program. Before the Connect, shares listed in China, called A-shares,
were only available to foreign institutional investors through certain
investment products called Qualified Foreign Institutional Investor
(QFII) funds and other such investment vehicles. Likewise, formal
channels of overseas investment by Chinese residents to the Qualified
Domestic Institutional Investor (QDII) schemes. The Connect program
allows a small number of securities listed on the Shanghai Stock Exchange
(SSE) to be traded by participants in Hong Kong and vice versa. Under
this pilot program, shares eligible to be traded through the Northbound
Trading Link (from Hong Kong to Shanghai) will comprise all the constituents
of the SSE 180 Index and SSE 380 Index, and shares of all SSE-listed
companies which have issued both A shares and H shares. Shares eligible
to be traded through the Southbound Trading Link (from Shanghai to
Hong Kong) comprise all the constituents of the Hang Seng Composite
Large Cap Index and Hang Seng Composite Mid Cap Index, and shares
of all companies listed on both SSE and Stock Exchange of Hong Kong
(SEHK). The initial expected date of the launch was towards the middle
of October. The delay also gives us a chance to study the reaction
of market participants on the announcement of the delay and the trends
in the run up to the final launch. 
\end{doublespace}
\begin{doublespace}

\subsection{Some Other Tidbits }
\begin{figure}[H]
\includegraphics{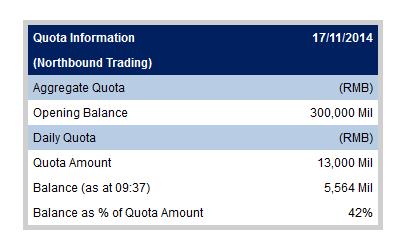}

\caption{First Seven Minutes of Trading\label{fig:First-Seven-Minutes}}

\end{figure}

\end{doublespace}
\begin{itemize}
\begin{doublespace}
\item Launched at 9:30 AM November 17, 2014. In the first seven minutes
(Figure \ref{fig:First-Seven-Minutes}), more than 50\% of the Quota
was used up (End-note \ref{enu:http://www.hkex.com.hk/eng/csm/c}).
\end{doublespace}
\end{itemize}
\begin{doublespace}

\end{doublespace}
\begin{itemize}
\begin{doublespace}
\item HK Exchange Total Market Cap \$3.1 trillion (Currently 6th globally),
Shanghai Exchange Total Market Cap \$2.4 trillion (7th) - Combined
Market Cap will put them in the top three. 
\item Currently QFII Quota (Includes other asset classes as well) around
US \$150 Billion. This is available only to large institutions 
\item 568 Securities in the Northbound (Hong Kong to Shanghai), 268 in the
Southbound (Shanghai to Hong Kong), 68 Dual Listed Securities 
\item Daily Northbound Quota is RMB 13 Billion (Around \$2 Billion USD),
Daily Southbound Quota is RMB 10.5 Billion (Less than \$2 Billion
USD) 
\item Market Cap (Approximate) on Hong Kong Connect Securities - \$105 Billion
USD 
\item Daily Average Daily Volume, ADV, (Approximate) on Hong Kong Connect
Securities - \$ 5 Billion USD 
\item Market Cap (Approximate) on Shanghai Connect Securities - \$53 Billion
USD 
\item Daily ADV (Approximate) on Shanghai Connect Securities - \$30 Billion
USD
\end{doublespace}
\end{itemize}
\begin{doublespace}

\subsection{Dual Listing Dynamics }
\end{doublespace}

\begin{doublespace}
There is again a massive amount of literature on dual listing and
its effect on stock returns. Serra (1999) investigates the effects
on stock returns of dual-listing on an international exchange across
a sample of emerging market securities. The results from the sample
they analyze confirm previous empirical findings that there are positive
abnormal returns before listing and a significant decline in expected
returns following the listing. Such a result is explained by the extent
of integration of capital markets, where integration is defined as
a situation where investors earn the same risk-adjusted expected return
on similar financial instruments in different markets. In a fully
integrated market, the price of risk would be the same in all markets
and this would be the compensation for the systemic risk factors that
occur globally. Where the extent of integration is less or where there
is segmentation, risk associated with local factors is priced into
the security returns, yielding different rewards to risk. Dual listing
or other forms of cross listing would mitigate segmentation by improving
risk sharing. The increased liquidity and investor awareness might
lower the required rate of return. Cross listing could also reduce
the cost of capital, acting as an incentive for firms. 

Shleifer and Vishny (1997) in their seminal work on the limits of
arbitrage point out that risk free arbitrage rarely exists in the
real world. Most arbitrageurs, who are specialized investors, managing
assets on behalf of others, face the possibility of interim liquidations
before the price disparity is restored, sometimes even when the arbitrage
opportunity is at its best, due to the short term losses that can
crop up from the diverging prices. This limitation on the capital
available for arbitrage could even amplify the arbitrage opportunity
by forcing the actions of investors against the direction of trading
that could potentially restore the price deviation and is compounded
due to any long run fundamental risk faced by the arbitrageurs. Gromb
and Vayanos (2010) emphasize costs that arbitrageurs could face, like
a) risk, both fundamental and non-fundamental; b) cost of short selling;
c) leverage and margin constraints; d) and constraints on equity capital.
Fundamental risk arises due to asset payoffs and non-fundamental risk
could be due to demand shocks generated by the set of investors who
are not solely looking to profit from the arbitrage opportunity, also
labeled “noise traders”. 

De Jong, Rosenthal and Van Dijk (2003, 2009) continue along this line
of inquiry by looking at price deviations across securities formed
as a result of mergers, in which both companies remain incorporated
independently. In contrast to securities listed on different exchanges
by the same company, the securities could be listed on the same exchange
in this case. To ensure there is no confusion in the terminology,
we refer to these as Siamese twins, as opposed to the listing of shares
in Hong Kong and China, which we refer to as dual listed. They show
that abnormal returns could persist even after accounting for systematic
risk, transaction costs and margin requirements. Market frictions
are controlled for by taking into account estimates of brokerage commissions,
bid-ask spreads, short rebates and capital requirements. The cost
of setting up an arbitrage position is taken as half of the bid-ask
spread, which lends some realism to the analysis yet remains questionable
of the real costs execution traders face while implementing portfolios
of large orders. Arbitrage activity could be further impeded due to
the volatility of returns from arbitrage and the high incidence of
negative returns, due to the uncertainty about the horizon at which
prices will converge and deviations from parity. Bedi, Richards and
Tennant (2003) highlight an interesting nuance that the pricing of
Siamese twins converges following the announcement of unification
via a raise in the price of the twin trading at a discount, while
confirming that a price divergence continues to exist for regular
listings on merged companies. 

Peng, Miao, and Chow (2008) argue that the extent of segmentation
between Hong Kong and China is high due to the restrictions imposed
on the mobility of capital. An interesting point to note is that almost
all dual listed companies have issued more H shares than A shares.
The relatively small supply of A shares could exacerbate the price
differential leading to the A share premium. Their investigation of
the price dynamics reveals that the A and H price differential is
stationary with a trend towards relative price convergence, where
the differential will not diverge persistently from a certain level,
as opposed to absolute price convergence or long term equalization,
where the price gap has a long term mean of zero. In addition to micro
factors, Fong, Wong and Yong (2008) consider macroeconomic factors
and find that Chinese currency appreciation expectations and monetary
expansion contributed to the A-H share price disparity by affecting
the prices of A shares, but their influence on the prices of H shares
was insignificant.
\end{doublespace}
\begin{doublespace}

\subsubsection{Main Difference from Dual Listing }
\end{doublespace}
\begin{itemize}
\begin{doublespace}
\item Comparing the Connect situation to a dual listed security, one main
difference would be the daily and aggregate quota limits on the amount
of securities, being bought and sold, that are part of this pilot
connect program. 
\item The other difference being that participants wishing to buy shares
in Shanghai need to go through their representatives or broker firms
in Hong Kong and vice versa.
\end{doublespace}
\end{itemize}
\begin{doublespace}

\subsection{Unintended Consequences}
\end{doublespace}

\begin{doublespace}
Any attempt at regulatory change is best exemplified by the story
of Sergey Bubka (End-note \ref{enu:Sergey-Nazarovich-Bubka}), the
Russian pole vault jumper, who broke the world record 35 times. Attempts
at regulatory change can be compared to taking the bar higher. In
this case, the intended effect of the change is to provide investors’
greater access to China markets without creating price distortions
and/or opportunities for abnormal profits. Despite all the uncertainty
(See Kashyap 2014a, 2014b), we can be certain of one thing, that the
market participants will find some way over the intended consequences,
prompting another round of rule revisions, or raising the bar, if
you will. (Kashyap 2015b) looks at a recent empirical example related
to trading costs where unintended consequences set in. Below we mention
the unknowns (or unintended consequences) that we know about (or can
anticipate). What about the unknowns that we don't know about (or
cannot even imagine). The only thing, we know about these unknown
unknowns are that, there must be a lot of them, hence the need for
us to be eternally vigilant, compelling all attempts at risk management
to make sure that the unexpected, even if it does happen, is contained
in the harm it can cause, while being cognizant that this is easier
said than done; a topic best saved for another time.
\end{doublespace}
\begin{doublespace}

\subsubsection{Uncertainty due to the Limited Quota }
\end{doublespace}
\begin{itemize}
\begin{doublespace}
\item Whenever there are limits imposed on the total amount of any good
and there is no explicit mechanism to distribute it, it can lead to
stock piling (in this case literally) and later distribution at a
profit. It remains to be seen, if the daily quota allowed in this
case is big enough to meet the demand or whether it would lead to
someone accumulating shares and parceling it out later. 
\item Imagine just one keg of beer, a few dozen college kids and no rules
as to how much beer each one gets!!! 
\item The imposed quota might exacerbate the possibility of arbitrage between
the connect shares across the two markets. 
\item Due to the quota there is some execution risk, which means if the
quota is filled just before an execution, the execution could get
rejected. This can happen despite the fact that the SEHK plans to
disseminate remaining quota balances every five seconds. 
\end{doublespace}
\end{itemize}
\begin{doublespace}

\subsection{A Recipe for the Skeptics}
\end{doublespace}

\begin{doublespace}
Among the key questions in the minds of many who wish to benefit from
this increased exposure to China, would be one key question \textbf{\textit{“Which
market would be better to trade to gain exposure to the same (or similar)
set of securities or sectors?” }}The answer to this question would
be determined by the implicit trading costs incurred on comparable
securities in Hong Kong and China.
\end{doublespace}
\begin{enumerate}
\begin{doublespace}
\item The main issue we run into, while doing any study on trading costs
in China, is that it is very hard to get a good sample of orders for
securities traded in China, before the connect was launched. 
\item Another related issue is that there are many ways to measure and estimate
implicit trading costs. The discussion on which ones are better can
prove to be very interesting, very long, and some would say, somewhat
inconclusive.
\end{doublespace}
\end{enumerate}
\begin{doublespace}
To solve both these problems, and perhaps make a case for even the
most hardened of skeptics amongst us, we design our study as follow: 
\end{doublespace}
\begin{itemize}
\begin{doublespace}
\item We run a set of simulations across dummy orders, with the same set
of parameters, and estimate implicit trading costs on dual listed
securities that trade both in China and Hong Kong. The estimated costs
on dummy orders, gives us a way of comparing similar orders trading
under similar market conditions in Hong Kong and Shanghai, and provides
a way to understand the trading cost trends in the two markets.

\end{doublespace}\begin{itemize}
\begin{doublespace}
\item We run time trend regressions across the market impact estimates,
which helps us understand which market is showing an increasing cost
trend. 
\item We do Welch t-tests (Welch 1947) on the estimated trading costs across
Hong Kong and Shanghai securities which gives us a way to assess which
time series is larger and hence instructive as to which market is
more expensive to trade. 
\end{doublespace}
\end{itemize}
\begin{doublespace}
\item We then look at the same set of metrics, estimated and actual trading
costs, across real orders on the Hong Kong securities in the dual
listed pair. This tells us how accurate our estimates are when compared
to actual trading costs.

\end{doublespace}\begin{itemize}
\begin{doublespace}
\item Lastly, we perform Mincer Zarnowitz regressions (Mincer and Zarnowitz
1969) to assess how good the forecasts are versus the observed values
on real orders, on which we have both the actual market impact costs
and the corresponding estimates. 
\end{doublespace}
\end{itemize}
\begin{doublespace}
\item We perform series of tests with different flavors, such as, considering:
the full sample; a sub sample two months before the event; taking
the simple average; taking the notional weighted average; excluding
high liquidity demanding orders; and aggregating costs across different
categories such as Side (Buy or Sell), Market Capitalization, Sector
and \%ADV demand of the orders.
\item Structuring the study in this way, helps us abstract away from many
of the nuances of how trading costs are measured and estimated. This
allows us to focus on the bigger puzzle of comparing the two markets.
\end{doublespace}
\end{itemize}
\begin{doublespace}
Despite this abstraction of some of the technical details, it is worthwhile
to have a brief sketch of our methodology to ensure that the reader
has no loss of continuity. The next section presents a synopsis of
our fundamentally different approach to Trading Cost Analysis. Kashyap
(2015c) is a complete development of this transaction cost model,
that incorporates stochastic dynamic programming based techniques
into the below formulation, under different laws of motion of the
security prices, starting with a benchmark scenario and extending
this to include multiple sources of uncertainty, liquidity constraints
due to volume curve shifts and relates trading costs to the spread.
We then move on to the numerical results, hoping to provide someone
looking to enter the Chinese Securities markets certain useful pieces
of information and to help them scale \textbf{\textit{“The Great Wall
of Chinese Securities Trading Costs”.}}
\end{doublespace}
\begin{doublespace}

\section{Trading Cost Measurement Methodology}
\end{doublespace}

\begin{doublespace}
The unique aspect of our approach to trading costs is a method of
splitting the overall move of the security price during the duration
of an order into two components (Collins and Fabozzi 1991; Treynor
1994; Yegerman and Gillula 2014). One component gives the costs of
trading that arise from the decision process that went into executing
that particular order, as captured by the price moves caused by the
executions that comprise that order. The other component gives the
costs of trading that arise due to the decision process of all the
other market participants during the time this particular order was
being filled. This second component is inferred, since it is not possible
to calculate it directly (at least with the present state of technology
and publicly available data) and it is difference between the overall
trading costs and the first component, which is the trading cost of
that order alone. The first and the second component arise due to
competing forces, one from the actions of a particular participant
and the other from the actions of everyone else that would be looking
to fulfill similar objectives. Naturally, it follows that each particular
participant can only influence to a greater degree the cost that arises
from his actions as compared to the actions of others, over which
he has lesser influence, but an understanding of the second component,
can help him plan and alter his actions, to counter any adversity
that might arise from the latter. Any good trader would do this intuitively
as an optimization process, that would minimize costs over two variables
direct impact and timing, the output of which recommends either slowing
down or speeding up his executions. With this measure, traders now
actually have a quantitative indicator to fine tune their decision
process. When we decompose the costs, it would be helpful to try and
understand how the two sub costs could vary as a proportion of the
total. The volatility in these two components, which would arise from
different sources (market conditions), would require different responses
and hence would affect the optimization problem mentioned above invoking
different sorts of handling and based on the situation, traders would
know which cost would be the more unpredictable one and hence focus
their efforts on minimizing the costs arising from that component.
Another popular way to decompose trading costs is into temporary and
permanent impact {[}See Almgren and Chriss (2001); Almgren (2003);
and Almgren, Thum, Hauptmann and Li (2005){]}. While the theory behind
this approach is extremely elegant and considers both linear and nonlinear
functions of the variables for estimating the impact, a practical
way to compute it requires measuring the price a certain interval
after the order. This interval is ambiguous and could lead to lower
accuracy while using this measure. 

We now introduce some terminology used throughout the discussion. 
\end{doublespace}
\begin{enumerate}
\begin{doublespace}
\item Total Slippage - The overall price move on the security during the
order duration. This is also a proxy for the implementation shortfall
(Perold 1988 and Treynor 1981). It is worth mentioning that there
are many similar metrics used by various practitioners and this concept
gets used in situations for which it is not the best suited (Yegerman
and Gillula 2014). While the usefulness of the Implementation Shortfall,
or slippage, as a measure to understand the price shortfalls that
can arise between constructing a portfolio and while implementing
it, is not to be debated, slippage need to be supplemented with more
granular metrics when used in situations where the effectiveness of
algorithms or the availability of liquidity need to be gauged. 
\item Market Impact (MI) - The price moves caused by the executions that
comprise the order under consideration. In short, the MI is a proxy
for the impact on the price from the liquidity demands of an order.
This metric is generally negative or zero since in most cases, the
best impact we can have is usually no impact. 
\item Market Timing - The price moves that happen due to the combined effect
of all the other market participants during the order duration. 
\item Market Impact Estimate (MIE) - This is the estimate of the Market
Impact, explained in point two above, based on recent market conditions.
The MIE calculation is the result of a simulation which considers
the number of executions required to fill an order and the price moves
encountered while filling this order, depending on the market micro-structure
as captured by the trading volume and the price probability distribution
including upticks and down-ticks, over the past few days. This simulation
can be controlled with certain parameters that dictate the liquidity
demanded on the order, the style of trading, order duration, market
conditions as reflected by start of trading and end of trading times.
In short, the MIE is an estimated proxy for the impact on the price
from the liquidity demands of an order. Such an approach holds the
philosophical viewpoint that making smaller predictions and considering
their combined effect would result in lesser variance as opposed to
making a large prediction; estimations done over a day as compared
to estimations over a month, say. A geometrical intuition would be
that fitting more lines (or curves) over a set of points would reduce
the overall error as compared to fitting lesser number of lines (or
curves) over the same set of points. When combining the results of
predictions, of course, we have to be mindful of the errors of errors,
which can get compounded and lead the results astray, and hence, empirical
tests need to be done to verify the suitability of such a technique
for the particular situation.
\item All these variables are measured in basis points to facilitate ease
of comparison and aggregation across different groups. It is possible
to measure these in cents per share and also in dollar value or other
currency terms. 
\item We start with equations, expressed in simple mathematical terms to
facilitate easier understanding, that govern the relationships between
the variables mentioned above. Next, we show two formulations of Market
Impact that can be fit into this framework, with a complete dynamic
programming approach available in (Kashyap 2015c).
\end{doublespace}
\end{enumerate}
\begin{doublespace}

\subsection{Market Impact Equations}
\end{doublespace}

\begin{doublespace}
Total Slippage = Market Impact + Market Timing 

\{Total Price Slippage = Your Price Impact + Price Impact From Everyone
Else (Price Drift)\} 

Market Impact Estimate = Market Impact Prediction = f (Execution Size,
Liquidity Demand) 

Execution Size = g(Execution Parameters, Market Conditions) 

Liquidity Demand = h(Execution Parameters, Market Conditions) 

Execution Parameters <->vector comprising (Order Size, Security, Side,
Trading Style, Timing Decisions) 

Market Conditions <-> vector comprising (Price Movement, Volume Changes,
Information Set)

Here, f, g, h are functions. We could impose concavity conditions
on these functions, but arguably, similar results are obtained by
assuming no such restrictions and fitting linear or non-linear regression
coefficients, which could be non-concave or even discontinuous allowing
for jumps in prices and volumes. The specific functional forms used
could vary across different groups of securities or even across individual
securities or even across different time periods for the same security.
The crucial aspect of any such estimation is the comparison with the
costs on real orders, as outlined earlier. Simpler modes are generally
more helpful in interpreting the results and for updating the model
parameters. Hamilton {[}1994{]} and Gujarati {[}1995{]} are classic
texts on econometrics methods and time series analysis that accentuate
the need for parsimonious models. 

The Auxiliary Information Set could be anything under our Sun or even
from under other heavenly objects. A useful variable to include would
be the blood pressure and heart rate time series of a representative
group of security traders.
\end{doublespace}
\begin{doublespace}

\subsubsection{Introducing our Innovation into the Implementation Shortfall}
\end{doublespace}

\begin{doublespace}
As a refresher, the total slippage or implementation shortfall is
derived below with the understanding that we need to use the Expectation
operator when we are working with estimates or future prices. (Kissell
2006) provides more details including the formula where the portfolio
may be partly executed. The list of symbols we use are,
\end{doublespace}
\begin{itemize}
\begin{doublespace}
\item $\bar{S}$, the total number of shares that need to be traded.
\item $T$, the total duration of trading.
\item $N$, the number of trading intervals.
\item $\tau=T/N$, the length of each trading interval. We assume the time
intervals are of the same duration, but this can be relaxed quite
easily. In continuous time, this becomes, $N\rightarrow\infty,\tau\rightarrow0$.
\item The time then becomes divided into discrete intervals, $t_{k}=k\tau,\;k=0,...,N$.
\item For simplicity, let time be measured in unit intervals giving, $t=1,2,...,T$.
\item $S_{t}$, the number of shares acquired in period $t$ at price $P_{t}$.
\item $P_{0}$ can be any reference price or benchmark used to measure the
slippage. It is generally taken to be the arrival price or the price
at which the portfolio manager would like to complete the purchase
of the portfolio.
\item Any trading trajectory, would look to formulate an optimal list of
total pending shares, $W_{1},...,W_{T+1}$. Here, $W_{t}$ is the
number of units that we still need to trade at time $t$. This would
mean, $W_{1}=\bar{S}$ and $W_{T+1}=0$ implies that $\bar{S}$ must
be executed by period $T$. Clearly, $\bar{S}=\underset{j=1}{\overset{T}{\sum}}S_{j}$.
This can equivalently be represented by the list of executions completed,
$S_{1},...,S_{T}$. Here, $W_{t}=W_{t-1}-S_{t-1}$ or $S_{t-1}=W_{t-1}-W_{t}$
is the number of units traded between times $t-1$ and $t$. $W_{t}$
and $S_{t}$ are related as below.
\begin{equation}
W_{t}=\bar{S}-\underset{j=1}{\overset{t-1}{\sum}}S_{j}=\underset{j=t}{\overset{T}{\sum}}S_{j}\qquad,t=1,...,T.
\end{equation}
Using the above notation, 
\end{doublespace}
\end{itemize}
\begin{doublespace}
\begin{equation}
\text{Paper Return}=\bar{S}P_{T}-\bar{S}P_{0}
\end{equation}
\begin{equation}
\text{Real Portfolio Return}=\bar{S}P_{T}-\left(\sum_{t=1}^{T}S_{t}P_{t}\right)
\end{equation}
\begin{align}
\text{Implementation Shortfall} & =\text{Paper Return}-\text{Real Portfolio Return}\\
 & =\left(\sum_{t=1}^{T}S_{t}P_{t}\right)-\bar{S}P_{0}
\end{align}
This can be written as,
\begin{align}
\text{Implementation Shortfall} & =\left(\sum_{t=1}^{T}S_{t}P_{t}\right)-\bar{S}P_{0}\\
 & =\left(\sum_{t=1}^{T}S_{t}P_{t}\right)-P_{0}\left(\sum_{t=1}^{T}S_{t}\right)\\
 & =S_{1}P_{1}+S_{2}P_{2}+...+S_{T}P_{T}-S_{1}P_{0}-S_{2}P_{0}-...-S_{T}P_{0}\\
 & =S_{1}\left(P_{1}-P_{0}\right)+S_{2}\left(P_{2}-P_{0}\right)+...+S_{T}\left(P_{T}-P_{0}\right)
\end{align}
\begin{align}
\text{Implementation Shortfall} & =S_{1}\left(P_{1}-P_{0}\right)\\
 & +S_{2}\left(P_{2}-P_{1}\right)+S_{2}\left(P_{1}-P_{0}\right)\\
 & +S_{3}\left(P_{3}-P_{2}\right)+S_{3}\left(P_{2}-P_{1}\right)+S_{3}\left(P_{1}-P_{0}\right)+\;...\;\\
 & +S_{T}\left(P_{T}-P_{T-1}\right)+S_{T}\left(P_{T-1}-P_{T-2}\right)+...+S_{T}\left(P_{1}-P_{0}\right)
\end{align}
The innovation we introduce would incorporate our earlier discussion
about breaking the total impact or slippage, Implementation Shortfall,
into the part from the participants own decision process, Market Impact,
and the part from the decision process of all other participants,
Market Timing. This Market Impact, would capture the actions of the
participant, since at each stage the penalty a participant incurs
should only be the price jump caused by their own trade and that is
what any participant can hope to minimize. A subtle point is that
the Market Impact portion need only be added up when new price levels
are established. If the price moves down and moves back up (after
having gone up once earlier and having been already counted in the
Impact), we need not consider the later moves in the Market Impact
(and hence implicitly left out from the Market Timing as well). This
alternate measure would only account for the net move in the prices
but would not show the full extent of aggressiveness and the push
and pull between market participants and hence is not considered here,
though it can be useful to know and can be easily incorporated while
running simulations. We discuss two formulations of our measure of
the Market Impact for a buy order, in the next two subsections. The
reason for calling them simple and complex will become apparent as
we continue the discussion.
\end{doublespace}
\begin{doublespace}

\subsubsection{Market Impact Simple Formulation}
\end{doublespace}

\begin{doublespace}
The simple market impact formulation does not consider the impact
of the new price level established on all the future trades that are
yet to be done. From a theoretical perspective it is useful to study
this since it provides a closed form solution and illustrates the
immense practical application of separating impact and timing. This
approach can be a useful aid in markets that are clearly not trending
and where the order size is relatively small compared to the overall
volume traded, ensuring that any new price level established does
not linger on for too long and prices gets reestablished due to the
trades of other participants. This property is akin to checking that
shocks to the system do not take long to dissipate and equilibrium
levels (or rather new pseudo equilibrium levels) are restored quickly.
Our measure of the Market Impact then becomes, 
\begin{equation}
\text{Market Impact}=\sum_{t=1}^{T}\left\{ \max\left[\left(P_{t}-P_{t-1}\right),0\right]S_{t}\right\} 
\end{equation}
The Market Timing is then given by,
\begin{align}
\text{\text{Market Timing}} & =\text{Implementation Shortfall}-\text{\text{Market Impact}}\\
 & =\left(\sum_{t=1}^{T}S_{t}P_{t}\right)-\bar{S}P_{0}-\sum_{t=1}^{T}\left\{ \max\left[\left(P_{t}-P_{t-1}\right),0\right]S_{t}\right\} 
\end{align}
For illustration, let us consider some examples,
\end{doublespace}
\begin{enumerate}
\begin{doublespace}
\item When all the successive price moves are above their corresponding
previous price, that is $\max\left[\left(P_{t}-P_{t-1}\right),0\right]=\left(P_{t}-P_{t-1}\right)$,
we have
\begin{align}
\text{Market Impact} & =\sum_{t=1}^{T}\left\{ \max\left[\left(P_{t}-P_{t-1}\right),0\right]S_{t}\right\} \\
 & =S_{1}\left(P_{1}-P_{0}\right)+S_{2}\left(P_{2}-P_{1}\right)+S_{3}\left(P_{3}-P_{2}\right)+\;...\;+S_{T}\left(P_{T}-P_{T-1}\right)
\end{align}
\begin{align}
\text{\text{Market Timing}} & =\text{Implementation Shortfall}-\text{\text{Market Impact}}\\
 & =\left(\sum_{t=1}^{T}S_{t}P_{t}\right)-\bar{S}P_{0}-S_{1}\left(P_{1}-P_{0}\right)-S_{2}\left(P_{2}-P_{1}\right)-S_{3}\left(P_{3}-P_{2}\right)-\;...\;-S_{T}\left(P_{T}-P_{T-1}\right)\\
 & =S_{1}P_{0}+S_{2}P_{1}+S_{3}P_{2}+\;...\;+S_{T}P_{T-1}-\bar{S}P_{0}\\
 & =S_{2}\left(P_{1}-P_{0}\right)+S_{3}\left(P_{2}-P_{0}\right)+\;...\;+S_{T}\left(P_{T-1}-P_{0}\right)
\end{align}
\item Some of the successive prices are below their corresponding previous
price, let us say, $\left(P_{2}<P_{1}\right)\text{ and }\left(P_{3}<P_{2}\right)$,
we have
\begin{align}
\text{Market Impact} & =\sum_{t=1}^{T}\left\{ \max\left[\left(P_{t}-P_{t-1}\right),0\right]S_{t}\right\} \\
 & =S_{1}\left(P_{1}-P_{0}\right)+S_{2}\left(0\right)+S_{3}\left(0\right)+\;...\;+S_{T}\left(P_{T}-P_{T-1}\right)
\end{align}
\begin{align}
\text{\text{Market Timing}} & =\text{Implementation Shortfall}-\text{\text{Market Impact}}\\
 & =\left(\sum_{t=1}^{T}S_{t}P_{t}\right)-\bar{S}P_{0}-S_{1}\left(P_{1}-P_{0}\right)-S_{2}\left(0\right)-S_{3}\left(0\right)-\;...\;-S_{T}\left(P_{T}-P_{T-1}\right)\\
 & =S_{2}P_{2}+S_{3}P_{3}+S_{1}P_{0}+S_{4}P_{3}+S_{5}P_{4}+\;...\;+S_{T}P_{T-1}-\bar{S}P_{0}\\
 & =S_{2}\left(P_{2}-P_{0}\right)+S_{3}\left(P_{3}-P_{0}\right)+S_{4}\left(P_{3}-P_{0}\right)+S_{5}\left(P_{4}-P_{0}\right)+\;...\;+S_{T}\left(P_{T-1}-P_{0}\right)
\end{align}
\end{doublespace}
\end{enumerate}
\begin{doublespace}

\subsubsection{\textcolor{black}{Market Impact Complex Formulation}}
\end{doublespace}

\begin{doublespace}
Another measure of the Market Impact can be formulated as below which
represents the idea that when a participant seeks liquidity and establishes
a new price level, all the pending shares or the unexecuted program
is affected by this new price level. This is a more realistic approach
since the action now will explicitly affect the shares that are not
yet executed. This measure can be written as, 
\begin{equation}
\text{Market Impact}=\sum_{t=1}^{T}\left\{ \max\left[\left(P_{t}-P_{t-1}\right),0\right]W_{t}\right\} 
\end{equation}
The Market Timing is then given by,
\begin{align}
\text{\text{Market Timing}} & =\text{Implementation Shortfall}-\text{\text{Market Impact}}\\
 & =\left(\sum_{t=1}^{T}S_{t}P_{t}\right)-\bar{S}P_{0}-\sum_{t=1}^{T}\left\{ \max\left[\left(P_{t}-P_{t-1}\right),0\right]W_{t}\right\} 
\end{align}
For illustration, let us consider some examples,
\end{doublespace}
\begin{enumerate}
\begin{doublespace}
\item When all the successive price moves are above their corresponding
previous price, that is $\max\left[\left(P_{t}-P_{t-1}\right),0\right]=\left(P_{t}-P_{t-1}\right)$,
we have
\begin{align}
\text{Market Impact} & =\sum_{t=1}^{T}\left\{ \max\left[\left(P_{t}-P_{t-1}\right),0\right]W_{t}\right\} \\
 & =W_{1}\left(P_{1}-P_{0}\right)+W_{2}\left(P_{2}-P_{1}\right)+W_{3}\left(P_{3}-P_{2}\right)+\;...\;+W_{T}\left(P_{T}-P_{T-1}\right)
\end{align}
\begin{align}
\text{\text{Market Timing}} & =\text{Implementation Shortfall}-\text{\text{Market Impact}}\\
 & =\left(\sum_{t=1}^{T}S_{t}P_{t}\right)-\bar{S}P_{0}-W_{1}\left(P_{1}-P_{0}\right)-W_{2}\left(P_{2}-P_{1}\right)-W_{3}\left(P_{3}-P_{2}\right)-\;...\;-W_{T}\left(P_{T}-P_{T-1}\right)\\
 & =\left[\sum_{t=1}^{T}\left(W_{t}-W_{t+1}\right)P_{t}\right]-W_{1}P_{0}-W_{1}\left(P_{1}-P_{0}\right)\\
 & -W_{2}\left(P_{2}-P_{1}\right)-W_{3}\left(P_{3}-P_{2}\right)-\;...\;-W_{T}\left(P_{T}-P_{T-1}\right)\\
 & =\left(W_{1}-W_{2}\right)P_{1}+\left(W_{2}-W_{3}\right)P_{2}+...+\left(W_{T}-W_{T+1}\right)P_{T}\\
 & -W_{1}P_{0}-W_{1}\left(P_{1}-P_{0}\right)-W_{2}\left(P_{2}-P_{1}\right)-W_{3}\left(P_{3}-P_{2}\right)-\;...\;-W_{T}\left(P_{T}-P_{T-1}\right)\\
 & =0
\end{align}
\item Some of the successive prices are below their corresponding previous
price, let us say, $\left(P_{2}<P_{1}\right)\text{ and }\left(P_{3}<P_{2}\right)$,
we have
\begin{align}
\text{Market Impact} & =\sum_{t=1}^{T}\left\{ \max\left[\left(P_{t}-P_{t-1}\right),0\right]W_{t}\right\} \\
 & =W_{1}\left(P_{1}-P_{0}\right)+W_{2}\left(0\right)+W_{3}\left(0\right)+\;...\;+W_{T}\left(P_{T}-P_{T-1}\right)
\end{align}
\begin{align}
\text{\text{Market Timing}} & =\text{Implementation Shortfall}-\text{\text{Market Impact}}\\
 & =\left(\sum_{t=1}^{T}S_{t}P_{t}\right)-\bar{S}P_{0}-W_{1}\left(P_{1}-P_{0}\right)-W_{2}\left(0\right)-W_{3}\left(0\right)-\;...\;-W_{T}\left(P_{T}-P_{T-1}\right)\\
 & =\left[\sum_{t=1}^{T}\left(W_{t}-W_{t+1}\right)P_{t}\right]-W_{1}P_{0}-W_{1}\left(P_{1}-P_{0}\right)\\
 & -W_{2}\left(0\right)-W_{3}\left(0\right)-\;...\;-W_{T}\left(P_{T}-P_{T-1}\right)\\
 & =\left(W_{1}-W_{2}\right)P_{1}+\left(W_{2}-W_{3}\right)P_{2}+...+\left(W_{T}-W_{T+1}\right)P_{T}\\
 & -W_{1}P_{0}-W_{1}\left(P_{1}-P_{0}\right)-W_{2}\left(0\right)-W_{3}\left(0\right)-\;...\;-W_{T}\left(P_{T}-P_{T-1}\right)\\
 & =-W_{2}P_{1}+W_{2}P_{2}-W_{3}P_{2}+W_{3}P_{3}\\
 & =W_{2}\left(P_{2}-P_{1}\right)+W_{3}\left(P_{3}-P_{2}\right)
\end{align}
\end{doublespace}
\end{enumerate}
\begin{doublespace}

\section{From Symbols to Numbers (From Modeling to Trading), Numerical Results}
\end{doublespace}

\begin{doublespace}
Adhering to a modified version of the old adage, “A picture is equal
to a thousand words or a million numbers (or pixels)”, we try to present,
where possible, the main empirical results as easy to read charts,
supplementing them with statistical tests and highlighting any major
trends with explanations. It is worth noting that majority of the
conclusions are fairly self-explanatory and some are possible to interpret
in different ways depending on the view one holds. The data-set and
the metrics are elaborated upon in the relevant sections below.
\end{doublespace}
\begin{doublespace}

\subsection{The Four Elements of the Empirical Study }
\end{doublespace}

\begin{doublespace}
We utilize a four pronged approach to understand the trading trends
due to the Hong Kong – Shanghai Connect. The four parts can be categorized
as follows 
\end{doublespace}
\begin{enumerate}
\begin{doublespace}
\item Volume: We look at the Volume Traded in the two markets across the
entire group of Connect securities. We also look at volume curves
across some single names in both Hong Kong and Shanghai across certain
key dates. The key dates we consider for the volume curves are

\end{doublespace}\begin{enumerate}
\begin{doublespace}
\item The start of the year, which also falls about three months before
the announcement of the program. This also captures any pre announcement
leakage of information. January 10, 2014 
\item The announcement date, April 10, 2014 
\item The initial expected launch date, October 27, 2014 
\item The actual launch date on November 17, 2014 
\end{doublespace}
\end{enumerate}
\begin{doublespace}
\item Price: The Price Convergence and Premium on dual listed securities
is analyzed in greater detail. This gives an indication of whether
prices are moving together or away from each other on similar securities
and hence helps shed some additional light on what we can expect from
trading costs. 
\item Market Impact or Implicit Trading Costs: We calculate the Market Impact
Estimate (MIE) on a sample of close to 500,000 dummy orders across
the dual listed securities with the same exact set of parameters.
We then look at the MIE and other trading cost metrics on close to
100,000 real orders across the dual listed Hong Kong securities. The
analysis time period for the simulation was from January 10, 2014
to November 14, 2014. The analysis time period for the real orders
was from January 10, 2014 to November 10, 2014. 
\item Auxiliary Order Level Metrics: We look at other useful metrics from
order level data including average trade size, average notional size,
percentage of spread paid, actual spread cost, order duration, number
of executions per minute and how the executions are dispersed over
the order interval. These auxiliary metrics are possible indicators
of changes in the trading strategies used over time.
\end{doublespace}
\end{enumerate}
\begin{doublespace}

\subsection{Stationary Tests on Volume and Price }
\end{doublespace}

\begin{doublespace}
We perform standard stationary tests on prices and volume on the overall
group of securities eligible for the Connect and also across just
the dual listed securities. We perform the Augmented Dickey-Fuller
(ADF) Test, the KPSS test and Phillips-Perron (PP) test (End-note
\ref{Statistical Significance Testing}; Dickey \& Fuller 1979; Bhargava
1986; Phillips \& Perron 1988; Kwiatkowski, Phillips, Schmidt, \&
Shin 1992; Greene 2003). The null hypothesis for the ADF and PP test
is that there is a unit root against the alternate that the series
is explosive or stationary. The KPSS null hypothesis is that the series
is level or trend stationary against the alternate that there is a
unit root. It is easily apparent that volumes are stationary and prices
are not across the entire group of the connect securities. The last
column, PP-stationary in Figure \ref{fig:Convergence-and-Stationary},
gives the count of securities with p-value less than the threshold
$\left(\alpha=0.05\right)$ yielding this inference. A similar result
holds when we look at volumes and prices across the dual listed pairs
of securities. 
\end{doublespace}
\begin{doublespace}

\subsection{Price Convergence of Dual Listed Securities}
\end{doublespace}

\begin{figure}[H]
\includegraphics[width=15cm]{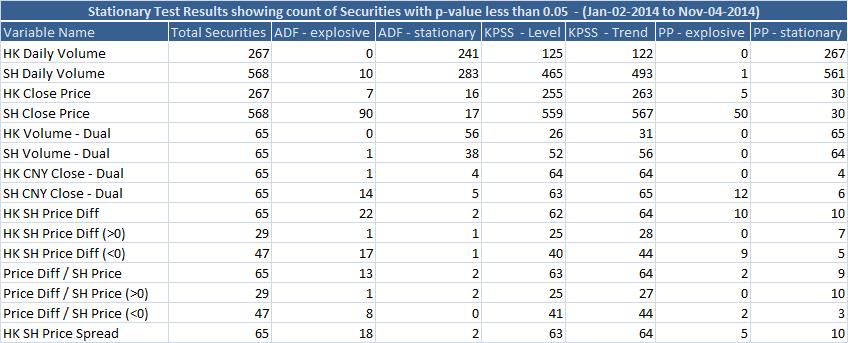}

\includegraphics[width=15cm]{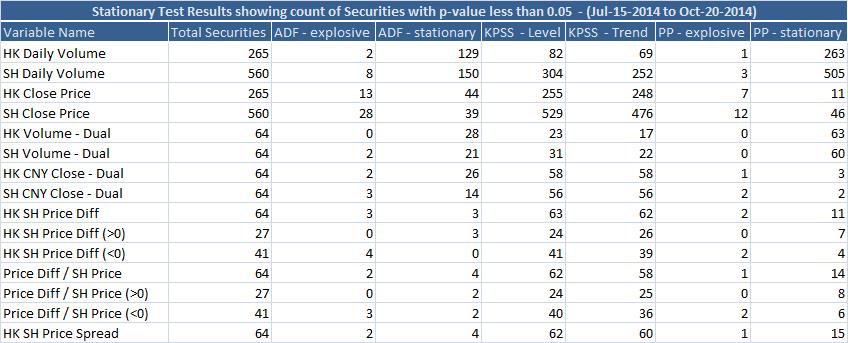}

\caption{Convergence and Stationary Test Results\label{fig:Convergence-and-Stationary}}

\end{figure}

\begin{doublespace}
We now zoom into the convergence of the prices of the dual listed
pair of securities. (Greasley \& Oxley 1997; Bernard \& Durlauf 1995;
1996) check for the convergence of economic time series based on unit
root tests. We apply similar methods to the price series of the dual
listed stocks to see if there are trends towards convergence. The
definition of convergence used implies that the below difference does
not converge if it contains a unit root.

\begin{align}
\underset{k\rightarrow\infty}{lim}E(y_{i,t+k}-y_{j,t+k} & \mid I_{t})=0\\
y_{i,t+k} & \leftrightarrow\text{Time\;\ Series\;\ for\;\ first\;\ security\;\ in\;\ the\;\ pair}\\
y_{j,t+k} & \leftrightarrow\text{Time\;\ Series\;\ for\;\ second\;\ security\;\ in\;\ the\;\ pair}\\
I_{t} & \leftrightarrow\text{Information\;\ Set}
\end{align}
We look at the below combinations of the dual listed prices while
checking for unit roots. When taking the difference, we always consider
the Hong Kong security as the first element. 
\end{doublespace}
\begin{enumerate}
\begin{doublespace}
\item The price difference between the dual listed pair denominated in the
Chinese currency. 
\item The price difference between the dual listed pair denominated in the
Chinese currency, where the difference is above zero. 
\item The price difference between the dual listed pair denominated in the
Chinese currency, where the difference is below zero. 
\item The price difference as a percentage of the price of the shanghai
security of the pair. 
\item The price difference as a percentage of the price of the shanghai
security of the pair, where the difference is above zero. 
\item The price difference as a percentage of the price of the shanghai
security of the pair, where the difference is below zero. 
\item We calculate the Hong Kong and Shanghai price spread and check whether
it is stationary. The spread is the error term when we run a regression
of one price against the other price in the security pair. 
\begin{align}
x_{it} & =\beta y_{it}+\varepsilon_{it}\\
x_{it} & \leftrightarrow\text{price\;\ of\;\ first\;\ security\;\ in\;\ the\;\ pair}\\
y_{it} & \leftrightarrow\text{price\;\ of\;\ second\;\ security\;\ in\;\ the\;\ pair}\\
\beta & \leftrightarrow\text{Linear\;\ Regression\;\ Coefficient}\\
\varepsilon_{it} & \leftrightarrow\text{error\;\ term}
\end{align}
\end{doublespace}
\end{enumerate}
\begin{doublespace}
The results are summarized in the figure \ref{fig:Convergence-and-Stationary}.
We find that there is a trend towards convergence among a subset of
the dual listed universe. It is worth noting that for the dual listed
securities, the sums of the security counts on the negative and positive
differences, do not add up to the security count on the aggregate,
indicating that some security pairs reverse the sign of their price
difference during this time period. The second table is the same set
of tests repeated over a shorter time frame. This gives us a chance
to check whether there is greater convergence in the dual listed price
pairs once participants have had long enough duration to react to
the initial announcement. The convergence is slightly higher for the
tests on the shorter time frame indicating that perhaps the delays
have contributed to the divergence again, though this is not significant.
(Su, Chong \& Yan 2007) find that there was convergence in the prices
after the launch of two policies, the QFII and the Closer Economic
Partnership Arrangement (CEPA). A point worth noting is that there
were less number of dual listed shares (less than half the number
now) at the time of their study. 

Some questions that bubble up to the surface are: 
\end{doublespace}
\begin{enumerate}
\begin{doublespace}
\item Do additional or newer dual listings bring in more divergence? With
the launch of the connect, it remains to be seen whether companies
would continue to prefer dual listings over listing on just one of
the exchanges, since, technically speaking, the two exchanges are
Connected (!).
\item Is convergence a temporary phenomenon?
\item Are the effects of direct trading related regulations (as the Connect)
stronger on convergence as compared to more indirect policy interventions
(as the QFII and CEPA)?
\end{doublespace}
\end{enumerate}
\begin{doublespace}

\end{doublespace}
\begin{doublespace}

\subsubsection{Price Premium}
\end{doublespace}

\begin{figure}[H]
\includegraphics[width=8cm,height=5cm]{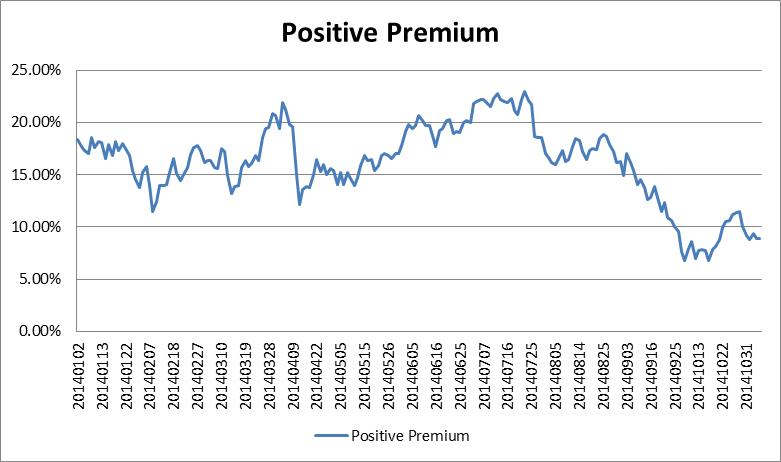}\includegraphics[width=8cm,height=5cm]{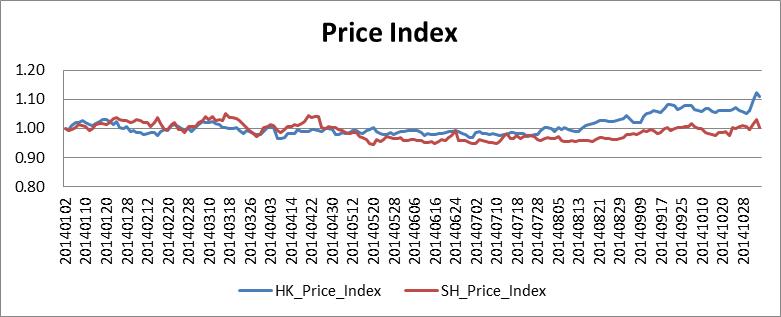}

\caption{Price Premium and Price Index\label{fig:Price-Premium-and}}
\end{figure}

\begin{doublespace}
The Positive Price Premium on dual listed securities has fallen by
more than 100\% (Figure \ref{fig:Price-Premium-and}). The positive
price premium is measured as the difference between the Hong Kong
Security Price and the Shanghai Security Price as a percentage of
the Shanghai price, when the Hong Kong Security Price is greater than
the Shanghai Price. While the below is the daily average across securities
with a positive premium, we see similar results for the aggregate
and also when we take a weighted average based on the volumes traded.
The median number of securities with the positive premium varies is
24 and it varies between 20 to 25 over the course of the analysis
time period. From the PP stationary test, there are more number of
securities here, as a proportion of the total that might have price
convergence. This explains the distinct change in the price premium
in this group, as compared to the negative premium and the aggregate
premium. The negative price premium is defined and treated similarly.
The median number of securities with the negative premium is 40 and
it varies between 40 to 45 over the course of the analysis time period.
We create a price index to show the overall price movement in all
the Hong Kong and Shanghai securities that are part of the connect,
weighted by market capitalization and the starting value set to one.
The Hong Kong price index is higher while the Shanghai price index
is not (Figure \ref{fig:Price-Premium-and}).

Additional graphs illustrating the total premium and how the premium
is distributed based on Market Cap and Turnover are given in Appendix
\ref{subsec:Price-Premium-Charts} (Figures \ref{fig:Total-and-Negative},
\ref{fig:Premium-vs-Combined}).

\end{doublespace}
\begin{doublespace}

\subsection{Hong Kong and Shanghai Traded Volume}
\end{doublespace}

\begin{figure}[H]
\includegraphics[width=15cm]{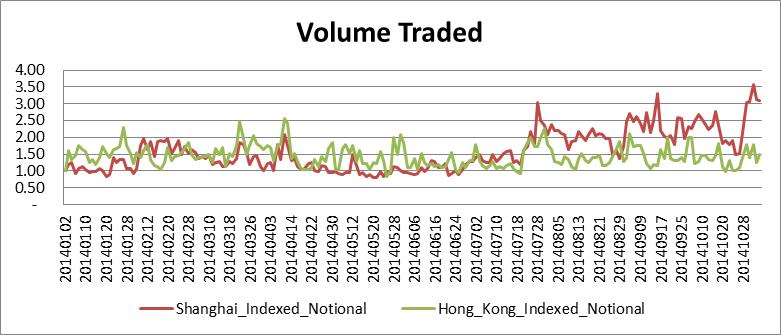}

\caption{Volume Traded\label{fig:Volume-Traded}}

\end{figure}

\begin{doublespace}
Volume Traded in Shanghai has gone up by more than 200\% (Figure \ref{fig:Volume-Traded}).
The volume is indexed to one at the start of the time period and the
effect of price increases have been removed to capture only the growth
in notional traded. Since volume is stationary from the earlier section,
we can conclude there is indeed a shift towards higher trading levels.

\end{doublespace}
\begin{doublespace}

\subsection{Simulated Trading Cost Comparison on Dual Listed Hong Kong and Shanghai
Securities}
\end{doublespace}

\begin{doublespace}
We look at how the trading cost estimates on dual listed securities
have changed over the months leading up to the Connect. The simulations
used have the same parameters for the Hong Kong security and for the
dual listed Shanghai security. We look at the full sample and also
slice it into various categories like Side, Market Capitalization,
Sector and \% ADV demand of the order. 
\end{doublespace}

We run time trend regressions of the sort below, where we aggregate
the market impact across different buckets by taking the notional
weighted average and for comparison purposes, we also consider another
version of these regressions just by taking the simple average. 
\begin{equation}
\overset{K}{\underset{i=1}{\sum}w_{i}}y_{it}=\beta_{0}+\beta_{1}t+\varepsilon_{t}
\end{equation}
Here, $w_{i}$ is the weight of an estimate $y_{it}$ in a particular
category being considered at time $t$ with a total of $K$ estimates
being aggregated in that bucket. 

\begin{doublespace}
We perform Welch-T tests across Hong Kong and Shanghai securities
for the simple average and notional weighted average trading cost
estimates over different buckets. The t statistic for this test, checks
the difference in the means of two time series and accounts for the
different variances.
\begin{equation}
t\quad=\quad\frac{\;\overline{X}_{1}-\overline{X}_{2}\;}{\sqrt{\;\frac{s_{1}^{2}}{N_{1}}\;+\;\frac{s_{2}^{2}}{N_{2}}\quad}}\,
\end{equation}
where $\overline{X}_{i}\text{, }s_{i}^{2}$ and $N_{i}$ are the sample
mean, sample variance and sample size, respectively.

We repeat the entire set of tests for the last two months before the
event. We perform another set of comparisons excluding liquidity demand
50\%+ ADV from the sample since the higher impact orders tend to be
larger and would skew the results. This would lead to better conclusions
because the number of real orders in these buckets tends to be small,
but we can still look at the changes in these higher ADV orders as
they will show up separately in the ADV categorization.

We see that the trading costs in Shanghai on an overall basis are
lower than Hong Kong till the days leading up to the connect, but
as we get closer to Connect, the costs in Shanghai become higher.
We summarize the comparison in figure \ref{fig:Simulation-Trading-Costs}
below. Additional graphs and results are given in Appendix \ref{subsec:Trading-Costs-Comparison}.
Figures \ref{fig:Simulation-Trading-Costs-1}, \ref{fig:Simulation-Trading-Costs-2}
and \ref{fig:Simulation-Trading-Costs-3} show trading cost trends
by liquidity demand, market cap and sectors.
\end{doublespace}

In the time trend regressions, we see more negative coefficients (also,
statistically significant) on the China sub-groups as compared to
the HK sub-groups. The results are only amplified when we consider
the full sample and the notional weighted average. Figures \ref{fig:Time-Trend-Regression},
\ref{fig:Time-Trend-Regression-1}, \ref{fig:Time-Trend-Regression-2},
\ref{fig:Time-Trend-Regression-3} and \ref{fig:Time-Trend-Regression-4}
report the time trend regressions (coefficients and corresponding
p-values are highlighted) when 50\% ADV orders are excluded by various
categories, Hong Kong securities, Shanghai securities, various categories
weighted by notional and for the last two months in the sample; Figure
\ref{fig:Time-Trend-Regression-5} is for the full sample including
the 50\%+ ADV orders.

In the Welch tests we see that the HK means are higher for the overall
sample, but in the last two months before the event the China means
are higher for the majority of the sub-groups. The last column in
Figures \ref{fig:Welch-T-Test}, \ref{fig:Welch-T-Test-1} and \ref{fig:Welch-T-Test-2}
shows the estimates of the mean values of HK and Shanghai, with less
than as the alternate hypothesis, when 50\% ADV orders are excluded,
for the last two months without 50\% ADV orders and for the full sample
including the 50\% ADV orders respectively.

\begin{doublespace}
\begin{figure}[H]
\includegraphics[width=16cm,height=7cm]{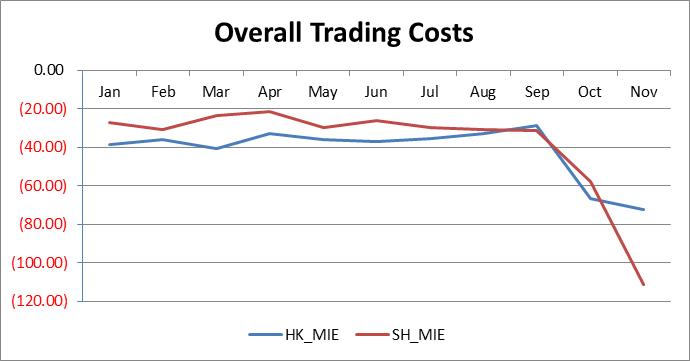}

\includegraphics[width=8cm,height=5cm]{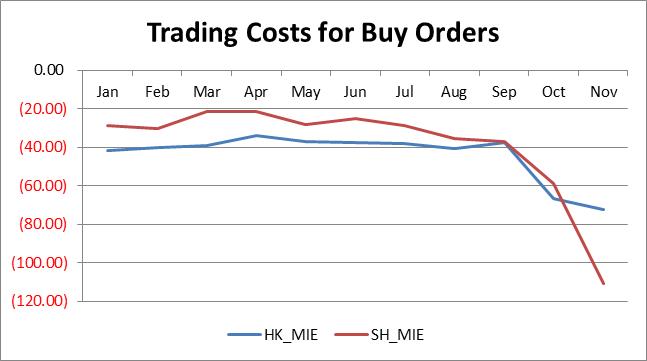}\includegraphics[width=8cm,height=5cm]{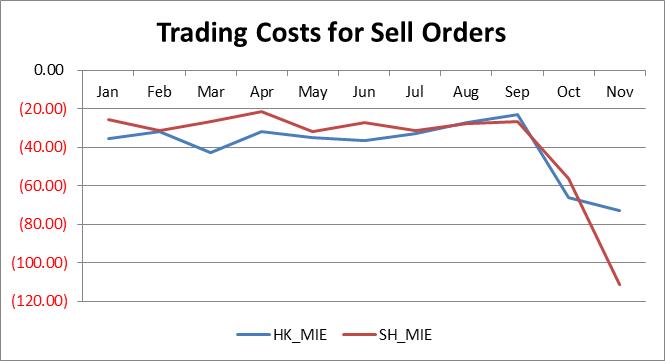}\caption{Simulation Trading Costs\label{fig:Simulation-Trading-Costs}}

\end{figure}

\end{doublespace}
\begin{doublespace}

\subsection{Estimated and Actual Costs across Real Orders on Dual Listed Hong
Kong Securities}
\end{doublespace}

\begin{doublespace}
When doing any study of trading costs, we need to face the realities
of high variance and extremely low co-efficient of variations. The
key is to extract the signal from the noise being mindful of the fact
that if we have a candle in the dark, our mission is accomplished.
To aid this effort at amplifying the signals, we filter out orders
that show zero market impact since they increase the noise without
adding any meaningful explanation. This is also practical from another
point since the orders with zero impact are traded very passively
and hence their inclusion would only reduce the contribution of orders
with meaningful impact towards any patterns we wish to uncover. To
differentiate this study from other impact studies that rely significantly
on price volatility, we run the test without including price volatility,
but use only a dummy variable to include the type of volatility environment.
We first run Mincer-Zarnowitz (MZ) type regressions of the type shown
below, between the actual impact and the corresponding estimate. 
\begin{equation}
y_{t}=\beta_{0}+\beta_{1}\hat{y_{t}}+\varepsilon_{t}
\end{equation}
Results of test of hypothesis (both joint and separate) on the estimated
coefficients using the F-test of significance for $\beta_{0}=0;\beta_{1}=1$
result in rejection; but with $\beta_{0}=\hat{\beta_{0}}\pm\triangle_{0};\beta_{1}=\hat{\beta_{1}}\pm\triangle_{1}$,
that is with small values around the estimated coefficients we get
high $p$-values (Hamilton 1994; Gujarati 1995; Verbeek 2008; End-notes
\ref{enu:F-Test-statistic:-The}, \ref{Statistical Significance Testing})
implying that the coefficients are significant but have a great deal
of sensitivity around their estimated values, an artifact of the high
noise environment.
\end{doublespace}

We run regressions on the full sample and also with trading costs
broken down into various categories similar to the ones we used in
the previous section. This allows the comparison of how accurate the
estimated costs are versus real costs and helps establish confidence
in our estimation methodology. The results (Appendix \ref{subsec:Comparison-of-Estimated})
show that the coefficients are non-zero and significant, indicating
a good level of forecasting prowess. Figures \ref{fig:Mincer-Zarnowitz-Regression}
and \ref{fig:Mincer-Zarnowitz-Regression-1} highlight the regression
coefficients and p-values for different groups and for HK securities.

To understand the upper limits of the predictive ability, we include
other variables and run secondary regressions. First we include category
variables. We try two flavors of specifications. One with the set
of category variables that we know before an order is traded (side,
capitalization, sector, liquidity demand). The other set would define
the environment when an order is being executed (expected price-momentum,
volume and volatility buckets). Other possible variables for the first
set are: arrival price, total number of shares, 90-day moving price
volatility for each security; for the second set, either as category
variables or explicit numerical forecasts, are: expectations regarding
notional traded, spread cost, number and size of executions, order
duration and security level price trend. This illustrates that specific
numerical forecasts of these second set of variables can enhance predictive
power, but even a judgment regarding which category might apply will
still be helpful towards improving performance. 
\begin{align}
y_{t} & =\beta_{0}+\beta_{1}\hat{y_{t}}+\overset{K}{\underset{i=2}{\sum}}\beta_{i}D_{i}+\varepsilon_{t}\\
\text{Here,} & D_{i}\text{ are dummy variables that define each of our sub groups with a total of }K\text{ subgroups.}
\end{align}
\begin{align}
y_{t} & =\beta_{0}+\beta_{1}\hat{y_{t}}+\overset{K}{\underset{i=2}{\sum}}\beta_{i}D_{i}+ln\left(Shares_{t}\right)+\left(Arrival\;Price_{t}\right)+\left(Moving\;Price\;Volatility_{\left\{ t-90,t\right\} }\right)\\
 & +\left(Other\;Variable\;Expectations_{t}\right)+\varepsilon_{t}
\end{align}

The evaluation of the forecasts still relies on the basic MZ regression.
The other specifications (Figure \ref{fig:Estimate-Regression-Results},
shows the results for two cuts of variables from among the many alternatives
tried) are merely to illustrate the increased explanatory power that
comes with our trading cost methodology. The full correlation matrix
is in Figure \ref{fig:Regression-Correlation-Matrix}. Figures (\ref{fig:Trading-Costs-on},
\ref{fig:Market-Impact-Costs}, \ref{fig:Market-Impact-Estimate},
\ref{fig:Market-Timing-Costs} and \ref{fig:Total-Slippage-Costs})
show time trends of all trading cost variables by side; Market Impact,
Market Impact Estimate, Market Timing and Total slippage by various
groups.
\begin{doublespace}

\subsection{Auxiliary Metrics on Real Orders}
\end{doublespace}

\begin{doublespace}
The price of liquidity, as measured by the spread, both in terms of
the actual value or in terms of percentage of the spread paid has
not changed drastically. The size of trades have increased both in
shares and notional terms, as has the duration over which orders are
traded. The velocity of trading as measured by the number of executions
per minute has decreased. This could be an indication of traders grabbing
bigger chunks of liquidity but more patiently since they are waiting
longer to fill the entire orders. Combining this inference with the
Volume Weighted Execution Time (VWET) we find that the trading is
still fairly evenly spread out over the duration of the order. VWET
indicates the extent to which executions are front loaded or back
loaded within the entire order duration. A value close to 50\% indicates
a fairly even distribution of executions or executions closer to the
front or to the back of the order duration. (Appendix \ref{subsec:Auxiliary-Metrics},
Figure \ref{fig:Auxiliary-Metrics-on})
\end{doublespace}
\begin{doublespace}

\subsection{Volume Curves for Select Hong Kong and Shanghai Names}
\end{doublespace}

\begin{doublespace}
Volume Curves across key dates leading up to the Connect are shown
for a select number of single names. It is fairly easy to infer that
the volumes traded increase significantly around announcement dates
and on the launch date of the Connect. We show the volume both as
percentage of day's total volume and in number of shares. (Appendix
\ref{subsec:Volume-Curves}, Figures \ref{fig:Volume-Curves-for}-\ref{fig:Volume-Curves-for-1})
\end{doublespace}
\begin{doublespace}

\section{Conclusions and Possibilities for Future Research }
\end{doublespace}

\begin{doublespace}
Adhering to a modified version of the old adage, “A picture is equal
to a thousand words or a million numbers (or pixels)”, we have presented,
where possible, some of the main empirical results as easy to read
graphs supplementing the analysis with statistical tests, explanations
and highlighting major observations. One conclusion that emerges is
that the trading costs in Shanghai which might have been cheaper compared
to Hong Kong might be becoming more expensive in the run up to the
Connect and perhaps even beyond. Contrary to what one would except,
given the increasing trading volume and converging price premium,
the divergence of trading costs stands out as an interesting effect
of the greater demand for liquidity on the northbound route. What
remains to be seen and analyzed in later studies is whether this increase
in trading costs is a temporary equilibrium due to the frenzy to gain
exposure to Chinese securities or whether this phenomenon will persist
once the two markets start becoming more and more tightly coupled.
Also of interest would be to see whether other regulatory interventions
in the financial markets will lead to such drastic changes to the
costs of trading.

In terms of methodology, different statistical procedures can be employed
in lieu of Welch type tests. (Welch 1938) considers in detail, tests
of hypothesis that the means of two normal populations are equal.
Yuen-Welch test (Yuen 1974); Brunner-Munzel test (Brunner and Munzel
2000); and Wilcoxon–Mann–Whitney test (Wilcoxon 1945; Mann and Whitney
1947) are some alternatives. (Fagerland and Sandvik 2009) compare
the performance of different tests for skewed distributions with unequal
variances. (Fagerland 2012) confirms that the Welch's t-test remains
robust for skewed distributions and large sample sizes. 

(Patton and Timmermann 2007a) study different tests of forecast optimality
and establish new testable properties that hold when the forecaster’s
loss function is unknown. (Patton and Timmermann 2007b) consider asymmetric
and nonlinear loss functions. (Elliott and Timmermann 2008) discuss
the central role of the loss function in helping determine the forecaster’s
objectives. They concede that the menu of forecasting methodologies
(none of which may coincide with the “true” model) has expanded vastly
over the last few decades. No single approach is currently dominant
and the choice of forecasting method is often dictated by the situation
at hand such as the forecast user’s particular needs, data availability,
and expertise in experimenting with different classes of models and
estimation methods. (Patton and Timmermann 2010) find that dispersion
among forecasters views is highest at long horizons. Our trading cost
methodology is based on the philosophy of short horizon forecasts,
and hence the simplicity of the MZ regression might be adequate for
our high variance environment. But we leave the door open to considering
variations to all the statistical methodologies we have employed,
which might show interesting results.

Once the actual connect program starts, we expect to have a significant
number of orders traded on securities listed in Shanghai through the
Connect. This will allow subsequent studies to do an actual comparison
on real orders of which market offers the better way to gain exposure
to similar securities from both a trading and also from a portfolio
construction perspective. Finally, as an afterthought we let the reader
ponder about what financial liberalizations means to the mode of governance
in a country. It might be an interesting study to look at other cases
where there have been significant changes to the extent of cross border
flows of capital and what effect it has had on the economy and the
overall well-being of the representative population (See Boyer and
Drache 1996; Kashyap 2015a; Quinn 2000; Simmons, Dobbin and Garrett
2008). A related question is the effect, the mode of governance and
other aspects of life in one country have on another country, once
they start linking up their financial markets.
\end{doublespace}
\begin{doublespace}

\subsection{Postscript}
\end{doublespace}

\begin{doublespace}
Since the launch of the connect, the trading volume has not been as
high as anticipated, though the program is claimed to be safe, stable
and a trend setter for similar partnerships being tabled globally,
despite some unresolved issues regarding beneficial and foreign ownership
rights, tax treatment on share gains and the custody of assets (End-notes
\ref{enu:http://en.xinfinance.com/html/Ma}, \ref{enu:http://www.hkex.com.hk/eng/marke}).
\end{doublespace}
\begin{doublespace}

\section{Acknowledgements and End-notes}
\end{doublespace}
\begin{enumerate}
\begin{doublespace}
\item \label{enu:The-author-would}The author would like to express his
gratitude to Brad Hunt, Henry Yegerman, Samuel Zou and Alex Gillula
at Markit; Patrick Lawlor, Joanna Wong and Eugene Kanvesky at CLSA,
for many inputs during the creation of this work. Dr. Yong Wang, Dr.
Isabel Yan, Dr. Vikas Kakkar, Dr. Fred Kwan, Dr. Srikant Marakani,
Dr. Costel Daniel Andonie, Dr. Jeff Hong, Dr. Guangwu Liu, Dr. Humphrey
Tung and Dr. Xu Han at the City University of Hong Kong provided advice
and more importantly encouragement to explore and where possible apply
cross disciplinary techniques. The views and opinions expressed in
this article, along with any mistakes, are mine alone and do not necessarily
reflect the official policy or position of either of my affiliations
or any other agency.
\item \label{enu:A-Review-of}In 2013, China’s real GDP grew by 7.7 \%,
the same as in 2012. Since 2010, the economic growth rate has declined
for four consecutive years (Fig. 2.1). Quite similar to the macroeconomic
trends and policy control mode adopted in 2012, from mid-2013, the
central government launched a series of fine-tuning measures to stabilize
the economic growth rate after experiencing sustained downward growth
during the first half of 2013. These measures inhibited the declining
economic trend in the third and fourth quarters and ensured that the
annual economic growth rate for the entire year matched that of the
previous year. However, the real annual growth rate of industrial
value added was only 9.7 \%, a decrease of 0.3 percentage points from
2012, and the lowest since 2009. A constant drop in the growth rate
of industrial value added reflects a slowdown in the real economy
(Center for Macroeconomic Research of Xiamen University 2015; A Review
of China’s Economy in 2013. Center for Macroeconomic Research, Xiamen
University. China’s Macroeconomic Outlook. Springer).
\end{doublespace}
\begin{doublespace}
\item \label{enu:http://www.hkex.com.hk/eng/csm/c}What is stock connect?
A unique collaboration between the Hong Kong, Shanghai and Shenzhen
Stock Exchanges, Stock Connect allows international and Mainland Chinese
investors to trade securities in each other's markets through the
trading and clearing facilities of their home exchange. \href{http://www.hkex.com.hk/eng/csm/chinaConnect.asp?LangCode=en}{What is stock connect?}
\item \label{enu:Sergey-Nazarovich-Bubka}Sergey Nazarovich Bubka (born
4 December 1963) is a Ukrainian former pole vaulter. He represented
the Soviet Union until its dissolution in 1991. Sergey has also beaten
his own record 14 times. He was the first pole vaulter to clear 6.0
metres and 6.10 metres. Bubka was twice named Athlete of the Year
by Track \& Field News and in 2012 was one of 24 athletes inducted
as inaugural members of the International Association of Athletics
Federations Hall of Fame. \href{https://en.wikipedia.org/wiki/Sergey_Bubka}{Sergey Bubka, Wikipedia Link}
\item \label{enu:http://en.xinfinance.com/html/Ma}The gong simultaneously
sounded in Shanghai Stock Exchange (SSE) and Hong Kong Exchanges and
Clearing Limited (HKEx) on Nov. 17 last year declared the official
launch of Shanghai-Hong Kong Stock Connect that links the capital
markets of the Chinese mainland and the world. Since then, direct
investment access to the stock markets in Shanghai and Hong Kong is
available. The stock market of the Chinese mainland is directly open
to global capital for the first time, while investors from the Chinese
mainland also start their way of global asset allocation. SSE and
HKEx report us the performance of the Shanghai-Hong Kong Stock Connect
at its anniversary. Though the overall transaction is not really hot,
but the program is “stable and safe” in operation and demonstrates
the whole world that such open model of capital market with joint
regulation from both sides, two-way access, closed operation and controllable
risk, pioneered by the Shanghai-Hong Kong Stock Connect, is completely
feasible. \href{http://en.xinfinance.com/html/Markets/SH_HK_Stock_Connect/2015/166963.shtml}{Shanghai-Hong Kong Stock Connect brings butterfly effect}
\end{doublespace}
\begin{doublespace}
\item \label{enu:http://www.hkex.com.hk/eng/marke}The establishment of
Shanghai-Hong Kong Stock Connect is a ground-breaking initiative to
both Mainland and Hong Kong as it has, for the first time, enabled
mutual market access by investors in the two markets through an orderly,
controllable and expandable channel. More importantly, this initiative
has paved the way for the opening up of the Mainland’s capital account
and helped promote the internationalization of Renminbi and development
of the Hong Kong’s capital market. \href{http://www.hkex.com.hk/eng/market/sec_tradinfra/chinaconnect/Documents/Investor_FAQ_En.pdf}{Shanghai-Hong Kong Stock Connect ... for Investors}
\end{doublespace}
\item \label{enu:F-Test-statistic:-The}F-Test statistic: The test for the
hypotheses,
\begin{align*}
\boldsymbol{H_{0}} & :\mu=k\\
\boldsymbol{H_{A}} & :\mu\neq k
\end{align*}
can be based on the F ratio,
\begin{align*}
F & =\frac{\text{Explained Variance}}{\text{Unexlpained Variance}}=\frac{\text{Between-Group Variability}}{\text{Within Group Variability}}\\
 & =\frac{n\left(\bar{Y}-k\right)^{2}}{S^{2}}\sim\left\{ \begin{array}{c}
F\left(1,n-1\text{ Degrees of Freedom}\right)\text{ If }\boldsymbol{H_{0}}\text{ is true}\\
\text{ Bigger}\qquad\qquad\qquad\qquad\text{ If }\boldsymbol{H_{A}}\text{ is true}
\end{array}\right.
\end{align*}
The p-value is the probability of such an 'extreme' value of the test
statistic when $\boldsymbol{H_{0}}$ is true. This is the upper tail
area of the $F$$\left(1,n-1\right)$ distribution. This p-value is
interpreted in exactly the same way as other p-values:
\begin{enumerate}
\item The smaller the p-value, the stronger the evidence that the null hypothesis
does not hold -{}- i.e. that is $\mu$ not equal to $k$. 
\item A large p-value (say 0.1 or higher) means that the data are consistent
with $\mu$ being equal to $k$.
\end{enumerate}
\begin{doublespace}
\href{http://www-ist.massey.ac.nz/dstirlin/CAST/CAST/HsimpleAnova/simpleAnova2.html}{P-Value for F-Test};
\href{https://en.wikipedia.org/wiki/F-test}{F-Test, Wikipedia Link}
\end{doublespace}
\item \label{Statistical Significance Testing}In general, if p-value is
less than a critical value then reject the null hypothesis. Less is
usually a p-value of 0.05 or lower. This would mean testing at the
5\% significance level.
\begin{enumerate}
\item For Stationary Tests, (ADF, PP, KPSS), the following are the null
and alternative hypotheses:
\item In ADF and PP, we specify the alternate hypothesis, stationary or
explosive. (Null is Unit Root). 
\item In KPSS, we specify the null hypothesis, Level or Trend Stationary.
(Alternative is Unit Root).
\end{enumerate}
\begin{doublespace}
\href{https://en.wikipedia.org/wiki/Augmented_Dickey–Fuller_test}{Augmented Dickey-Fuller Test, Wikipedia Link};
\href{https://en.wikipedia.org/wiki/Phillips–Perron_test}{Phillips–Perron Test, Wikipedia Link};
\href{https://en.wikipedia.org/wiki/KPSS_test}{KPSS-Test, Wikipedia Link}
\end{doublespace}
\end{enumerate}

\section{References}
\begin{enumerate}
\begin{doublespace}
\item Almgren, R., \& Chriss, N. (2001). Optimal execution of portfolio
transactions. Journal of Risk, 3, 5-40.
\item Almgren, R. F. (2003). Optimal execution with nonlinear impact functions
and trading-enhanced risk. Applied mathematical finance, 10(1), 1-18.
\item Almgren, R., Thum, C., Hauptmann, E., \& Li, H. (2005). Direct estimation
of equity market impact. Risk, 18, 5752.
\item Beck, T., \& Levine, R. (2004). Stock markets, banks, and growth:
Panel evidence. Journal of Banking \& Finance, 28(3), 423-442.
\item Bedi, J., Richards, A. J., \& Tennant, P. (2003). The characteristics
and trading behavior of dual-listed companies. Reserve Bank of Australia
Research Discussion Paper, (2003-06).
\item Bekaert, G., Harvey, C. R., \& Lundblad, C. (2005). Does financial
liberalization spur growth?. Journal of Financial economics, 77(1),
3-55.
\item Bekaert, G., Harvey, C. R., \& Lundblad, C. (2007). Liquidity and
expected returns: Lessons from emerging markets. Review of Financial
Studies, 20(6), 1783-1831.
\item Bernard, A. B., \& Durlauf, S. N. (1995). Convergence in international
output. Journal of applied econometrics, 10(2), 97-108.
\item Bernard, A. B., \& Durlauf, S. N. (1996). Interpreting tests of the
convergence hypothesis. Journal of econometrics, 71(1), 161-173.
\end{doublespace}
\item Bhargava, A. (1986). On the theory of testing for unit roots in observed
time series. The Review of Economic Studies, 53(3), 369-384.
\begin{doublespace}
\item Brunner, E., \& Munzel, U. (2000). The nonparametric Behrens‐Fisher
problem: Asymptotic theory and a small‐sample approximation. Biometrical
Journal, 42(1), 17-25.
\item Boyer, R. \& Drache, D. (Eds.). (1996). States against markets: the
limits of globalization. York University. University of Toronto. Innis
College (Toronto). London: Routledge.
\end{doublespace}
\item Center for Macroeconomic Research of Xiamen University. (2015). A
Review of China’s Economy in 2013. In: China’s Macroeconomic Outlook.
Current Chinese Economic Report Series. pp 7-18. Springer, Berlin,
Heidelberg.
\begin{doublespace}
\item Collins, B. M., \& Fabozzi, F. J. (1991). A methodology for measuring
transaction costs. Financial Analysts Journal, 47(2), 27-36.
\item Deeg, R., \& O'Sullivan, M. A. (2009). The political economy of global
finance capital. World Politics, 61(04), 731-763.
\item De Jong, A., Rosenthal, L., \& Van Dijk, M. A. (2003). The limits
of arbitrage: evidence from dual-listed companies. Erasmus University
working paper.
\item De Jong, A., Rosenthal, L., \& Van Dijk, M. A. (2009). The Risk and
Return of Arbitrage in Dual-Listed Companies{*}. Review of Finance,
13(3), 495-520.
\end{doublespace}
\item Dickey, D. A., \& Fuller, W. A. (1979). Distribution of the estimators
for autoregressive time series with a unit root. Journal of the American
statistical association, 74(366a), 427-431.
\begin{doublespace}
\item Elliott, G., \& Timmermann, A. (2008). Economic Forecasting. Journal
of Economic Literature, 46(1), 3-56.
\item Epstein, G. A. (Ed.). (2005). Financialization and the world economy.
Edward Elgar Publishing. Cheltenham, UK.
\item Fagerland, M. W., \& Sandvik, L. (2009). Performance of five two-sample
location tests for skewed distributions with unequal variances. Contemporary
clinical trials, 30(5), 490-496.
\item Fagerland, M. W. (2012). t-tests, non-parametric tests, and large
studies—a paradox of statistical practice?. BMC medical research methodology,
12(1), 1.
\item Fong, T., Wong, A., \& Yong, I. (2008). Share price disparity in Chinese
stock markets. Macroeconomic Linkages between Hong Kong and Mainland
China.
\item Greasley, D., \& Oxley, L. (1997). Time-series based tests of the
convergence hypothesis: some positive results. Economics Letters,
56(2), 143-147.
\end{doublespace}
\item Greene, W. H. (2003). Econometric analysis. Pearson Education India.
\begin{doublespace}
\item Gromb, D., \& Vayanos, D. (2010). Limits of Arbitrage. Annual Review
of Financial Economics, 2(1), 251-275.
\item Gujarati, D. N. (1995). Basic econometrics, 3rd. International Edition.
\item Hamilton, J. D. (1994). Time series analysis (Vol. 2). Princeton university
press.
\item Henry, P. B. (2000). Do stock market liberalizations cause investment
booms?. Journal of Financial economics, 58(1), 301-334.
\item Kashyap, R. (2014a). Dynamic Multi-Factor Bid–Offer Adjustment Model.
The Journal of Trading, 9(3), 42-55.
\item Kashyap, R. (2014b). The Circle of Investment. International Journal
of Economics and Finance, 6(5), 244-263.
\item Kashyap, R. (2015a). Financial Services, Economic Growth and Well-Being:
A Four Pronged Study. Indian Journal of Finance, 9(1), 9-22.
\item Kashyap, R. (2015b). A Tale of Two Consequences. The Journal of Trading,
10(4), 51-95.
\item Kashyap, R. (2015c). David vs Goliath (You against the Markets), A
Dynamic Programming Approach to Separate the Impact and Timing of
Trading Costs. Working Paper.
\item Kissell, R. (2006). The expanded implementation shortfall: Understanding
transaction cost components. The Journal of Trading, 1(3), 6-16.
\end{doublespace}
\item Kwiatkowski, D., Phillips, P. C., Schmidt, P., \& Shin, Y. (1992).
Testing the null hypothesis of stationarity against the alternative
of a unit root: How sure are we that economic time series have a unit
root?. Journal of econometrics, 54(1-3), 159-178.
\begin{doublespace}
\item Levine, R. (2001). International financial liberalization and economic
growth. Review of International economics, 9(4), 688-702.
\item Levine, R., \& Zervos, S. (1996). Stock market development and long-run
growth. The World Bank Economic Review, 10(2), 323-339.
\item Levine, R., \& Zervos, S. (1998a) Stock Markets, Banks, and Economic
Growth. American economic review. 88(3), 537-558.
\item Levine, R., \& Zervos, S. (1998b). Capital control liberalization
and stock market development. World Development, 26(7), 1169-1183.
\item Mann, H. B., \& Whitney, D. R. (1947). On a test of whether one of
two random variables is stochastically larger than the other. The
annals of mathematical statistics, 50-60.
\item Mincer, J. A., \& Zarnowitz, V. (1969). The evaluation of economic
forecasts. In Economic Forecasts and Expectations: Analysis of Forecasting
Behavior and Performance (pp. 3-46). NBER.
\item Patton, A. J., \& Timmermann, A. (2007a). Testing Forecast Optimality
Under Unknown Loss. Journal of the American Statistical Association,
102(480), 1172-1184.
\item Patton, A. J., \& Timmermann, A. (2007b). Properties of optimal forecasts
under asymmetric loss and nonlinearity. Journal of Econometrics, 140(2),
884-918.
\item Patton, A. J., \& Timmermann, A. (2010). Why do forecasters disagree?
Lessons from the term structure of cross-sectional dispersion. Journal
of Monetary Economics, 57(7), 803-820.
\item Peng, W., Miao, H., \& Chow, N. (2008). Price convergence between
dual-listed A and H shares. Macroeconomic Linkages between Hong Kong
and Mainland China, 295-315.
\item Perold, A. F. (1988). The implementation shortfall: Paper versus reality.
The Journal of Portfolio Management, 14(3), 4-9.
\end{doublespace}
\item Phillips, P. C., \& Perron, P. (1988). Testing for a unit root in
time series regression. Biometrika, 75(2), 335-346.
\begin{doublespace}
\item Quinn, D. P. (2000). Democracy and international financial liberalization.
McDonough School of Business, Georgetown University.
\item Shleifer, A., \& Vishny, R. W. (1997). The limits of arbitrage. The
Journal of Finance, 52(1), 35-55.
\item Simmons, B. A., Dobbin, F., \& Garrett, G. (Eds.). (2008). The global
diffusion of markets and democracy (pp. 319-332). Cambridge: Cambridge
University Press.
\item Serra, A. P. (1999). Dual‐listings on international exchanges: the
case of emerging markets’ stocks. European Financial Management, 5(2),
165-202.
\item Su, Q., Chong, T. T. L., \& Yan, I. K. M. (2007). On the convergence
of the Chinese and Hong Kong stock markets: a cointegration analysis
of the A and H shares. Applied Financial Economics, 17(16), 1349-1357.
\item Treynor, J. L. (1981). What does it take to win the trading game?.
Financial Analysts Journal, 37(1), 55-60.
\item Treynor, J. L. (1994). The invisible costs of trading. The Journal
of Portfolio Management, 21(1), 71-78.
\end{doublespace}
\item Verbeek, M. (2008). A guide to modern econometrics. John Wiley \&
Sons.
\begin{doublespace}
\item Welch, B. L. (1938). The significance of the difference between two
means when the population variances are unequal. Biometrika, 29(3-4),
350-362.
\item Welch, B. L. (1947). The generalization of student's' problem when
several different population variances are involved. Biometrika, 34(1-2),
28-35.
\item Wilcoxon, F. (1945). Individual comparisons by ranking methods. Biometrics
bulletin, 1(6), 80-83.
\item Yegerman, H. \& Gillula, A. (2014). The Use and Abuse of Implementation
Shortfall. Markit Working Paper.
\item Yuen, K. K. (1974). The two-sample trimmed t for unequal population
variances. Biometrika, 61(1), 165-170.
\end{doublespace}
\end{enumerate}
\begin{doublespace}

\section{Appendix}
\end{doublespace}
\begin{doublespace}

\subsection{Price Premium Charts\label{subsec:Price-Premium-Charts}}
\end{doublespace}

\begin{doublespace}
The Total Premium is the average difference in price between Hong
Kong price and Shanghai price expressed as a percentage of the Shanghai
Price. The Combined Market Capitalization is the sum of the Market
Cap of the Hong Kong and Shanghai Security expressed in USD Billions.
The Combined Turnover is the sum of the turnover of the Hong Kong
and Shanghai Security, with both using the price of Shanghai security
expressed in CNY Billions. 

\begin{figure}[H]
\includegraphics[width=8cm,height=5cm]{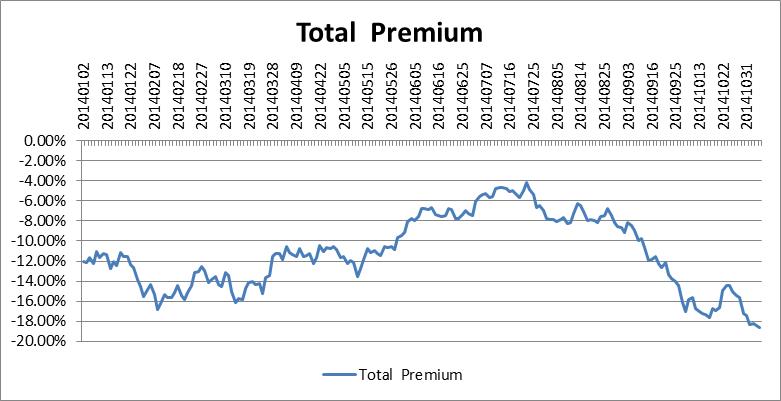}\includegraphics[width=8cm,height=5cm]{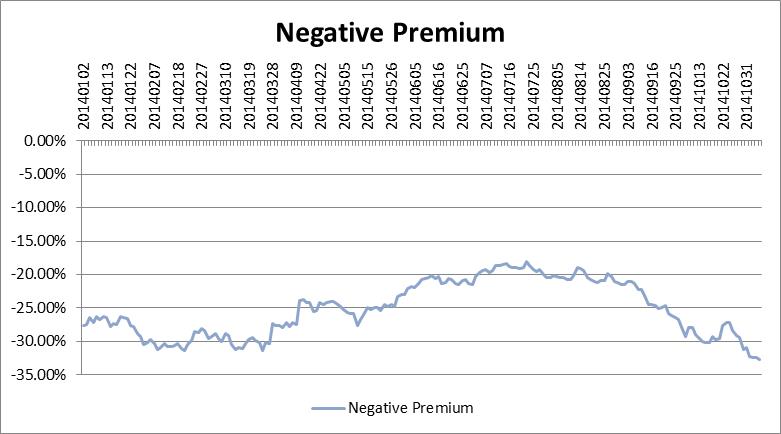}

\caption{Total and Negative Premium\label{fig:Total-and-Negative}}
\end{figure}

\begin{figure}[H]
\includegraphics[width=8cm,height=5cm]{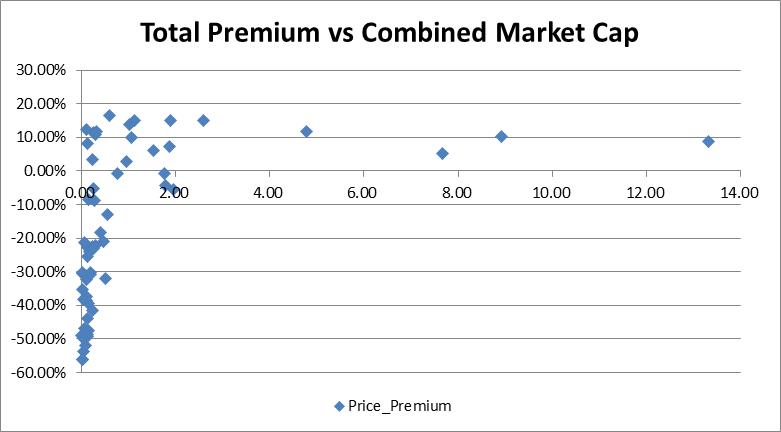}\includegraphics[width=8cm,height=5cm]{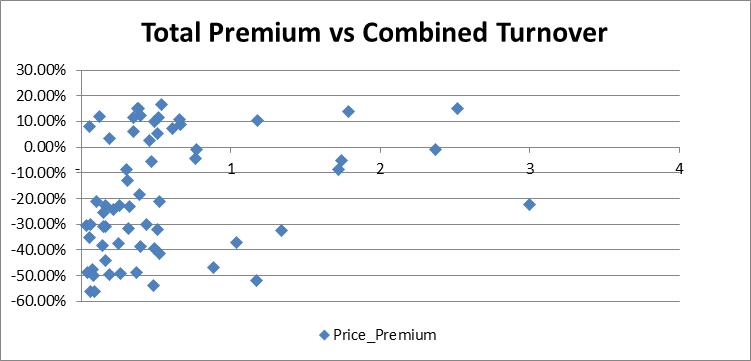}

\caption{Premium vs Combined Market Cap and Turnover\label{fig:Premium-vs-Combined}}
\end{figure}

\end{doublespace}
\begin{doublespace}

\subsection{Trading Cost Comparisons between HK and China using Simulations\label{subsec:Trading-Costs-Comparison}}
\end{doublespace}

\begin{doublespace}
\begin{figure}[H]
\includegraphics[width=18cm]{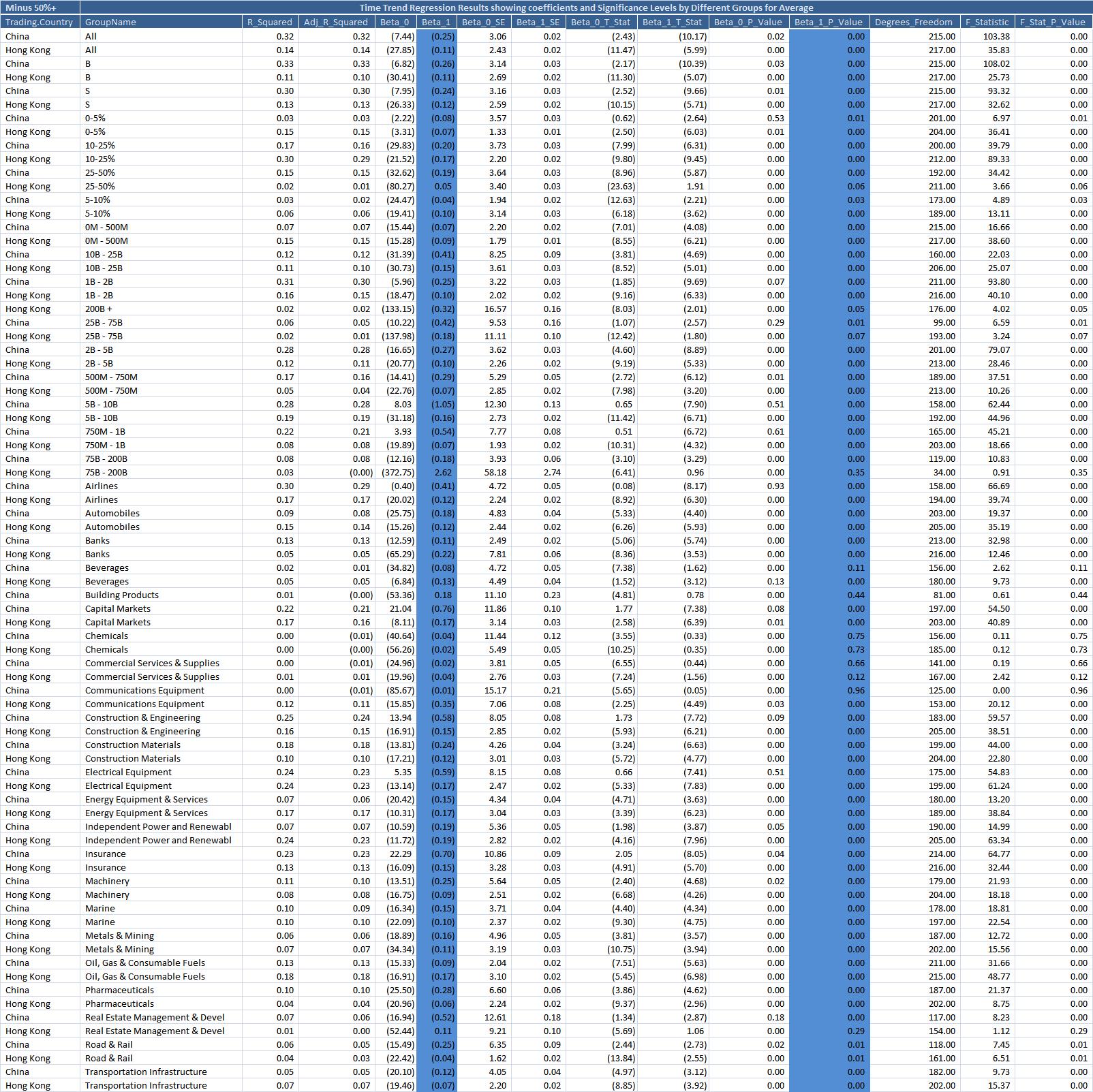}

\caption{Time Trend Regression - Without 50\%+ ADV\label{fig:Time-Trend-Regression}}
\end{figure}

\begin{figure}[H]
\includegraphics[width=18cm]{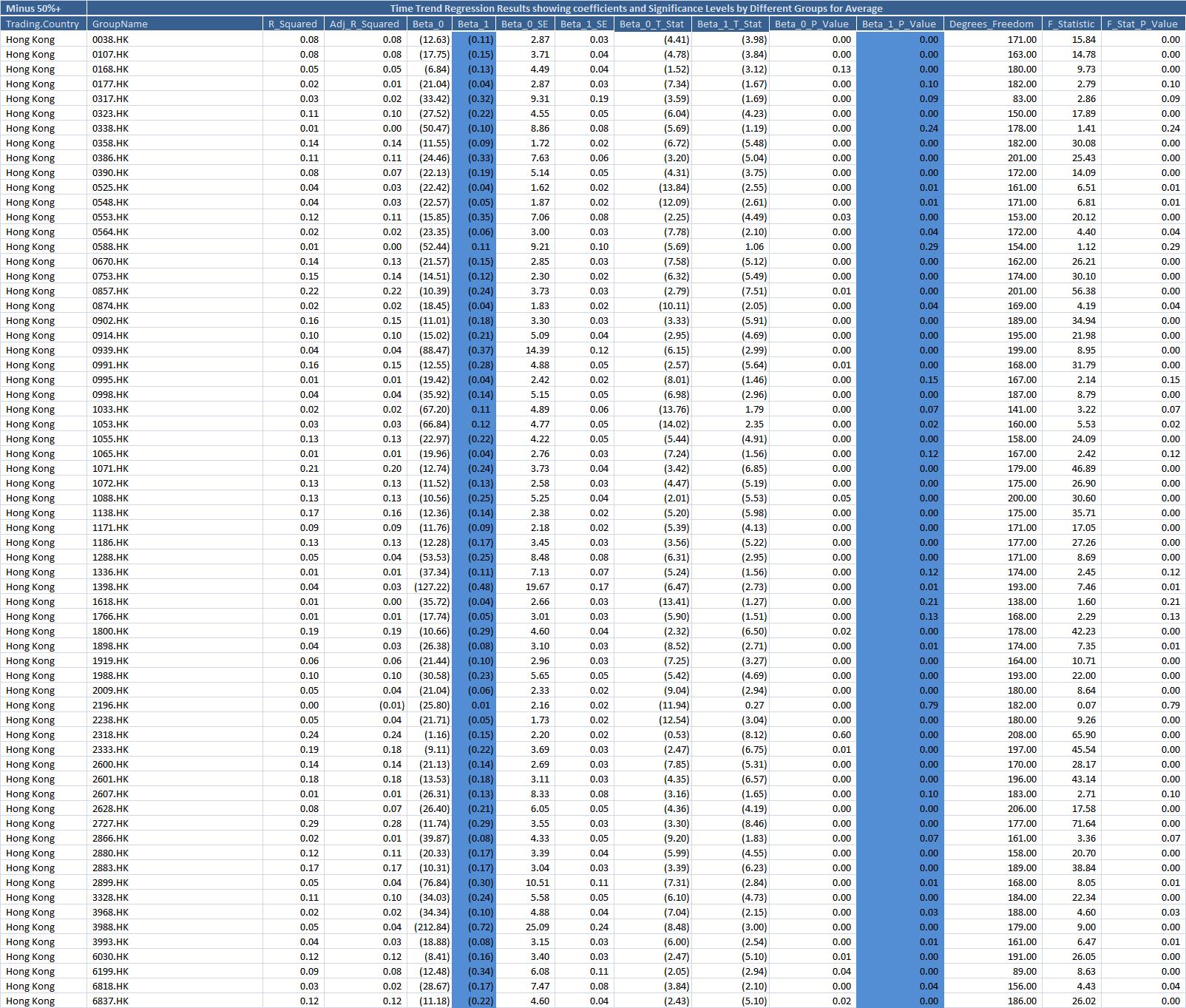}

\caption{Time Trend Regression - Without 50\%+ ADV HK Securities\label{fig:Time-Trend-Regression-1}}
\end{figure}

\begin{figure}[H]
\includegraphics[width=18cm]{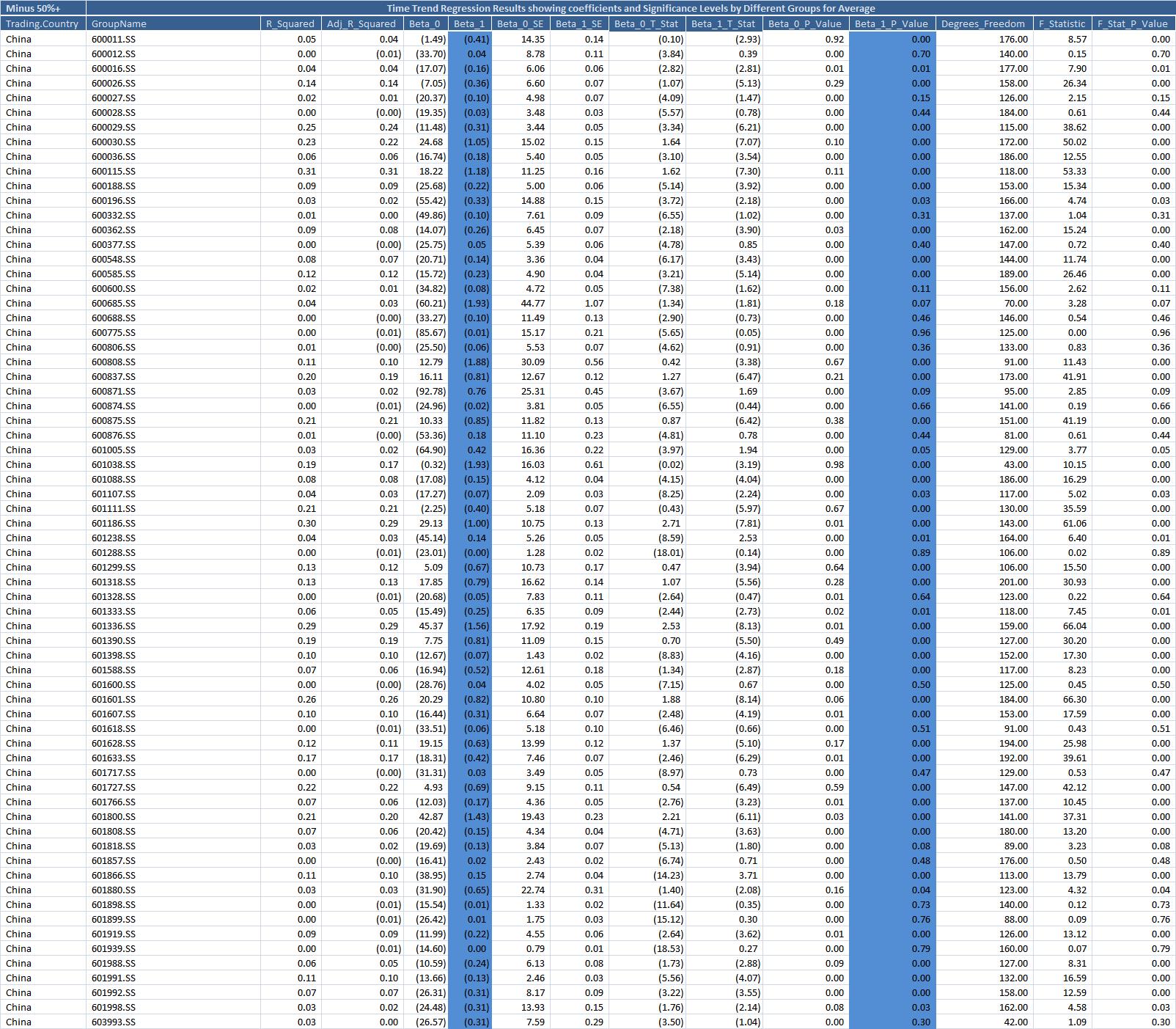}

\caption{Time Trend Regression - Without 50\%+ ADV SH Securities\label{fig:Time-Trend-Regression-2}}
\end{figure}

\begin{figure}[H]
\includegraphics[width=18cm]{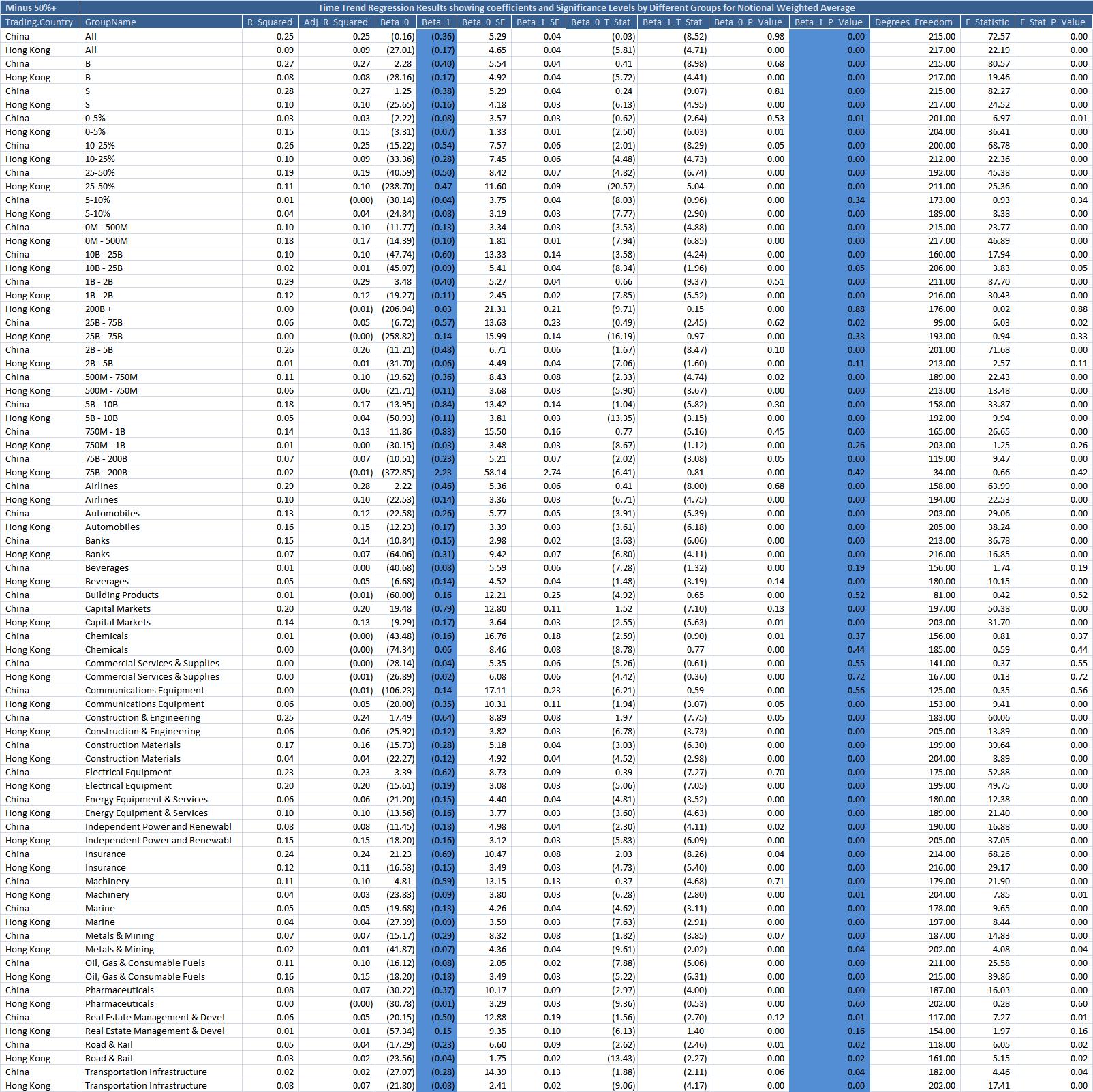}

\caption{Time Trend Regression - Without 50\%+ ADV Notional Weighted\label{fig:Time-Trend-Regression-3}}
\end{figure}

\begin{figure}[H]
\includegraphics[width=18cm]{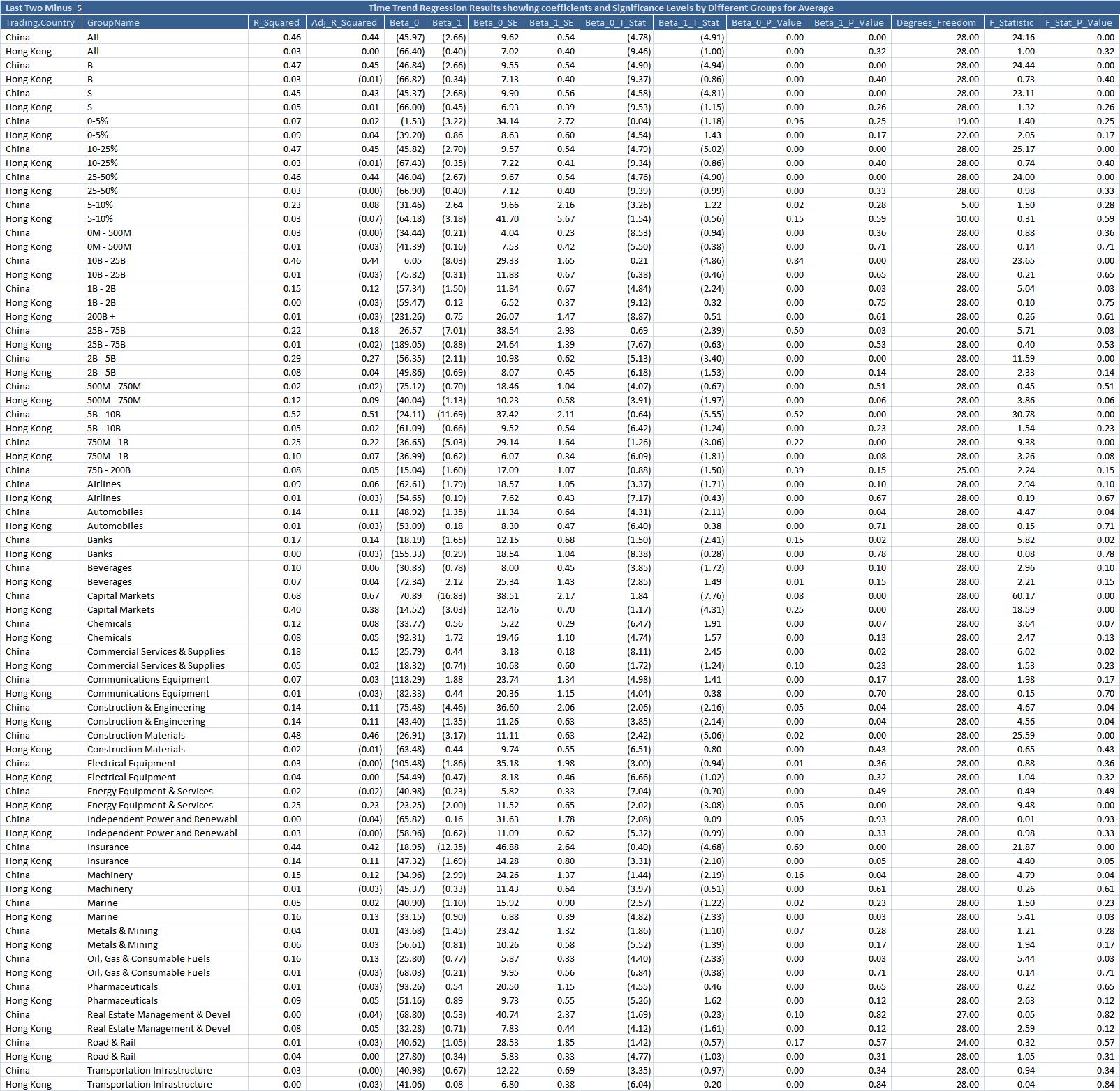}

\caption{Time Trend Regression - Without 50\%+ ADV Last Two Months\label{fig:Time-Trend-Regression-4}}
\end{figure}

\begin{figure}[H]
\includegraphics[width=18cm]{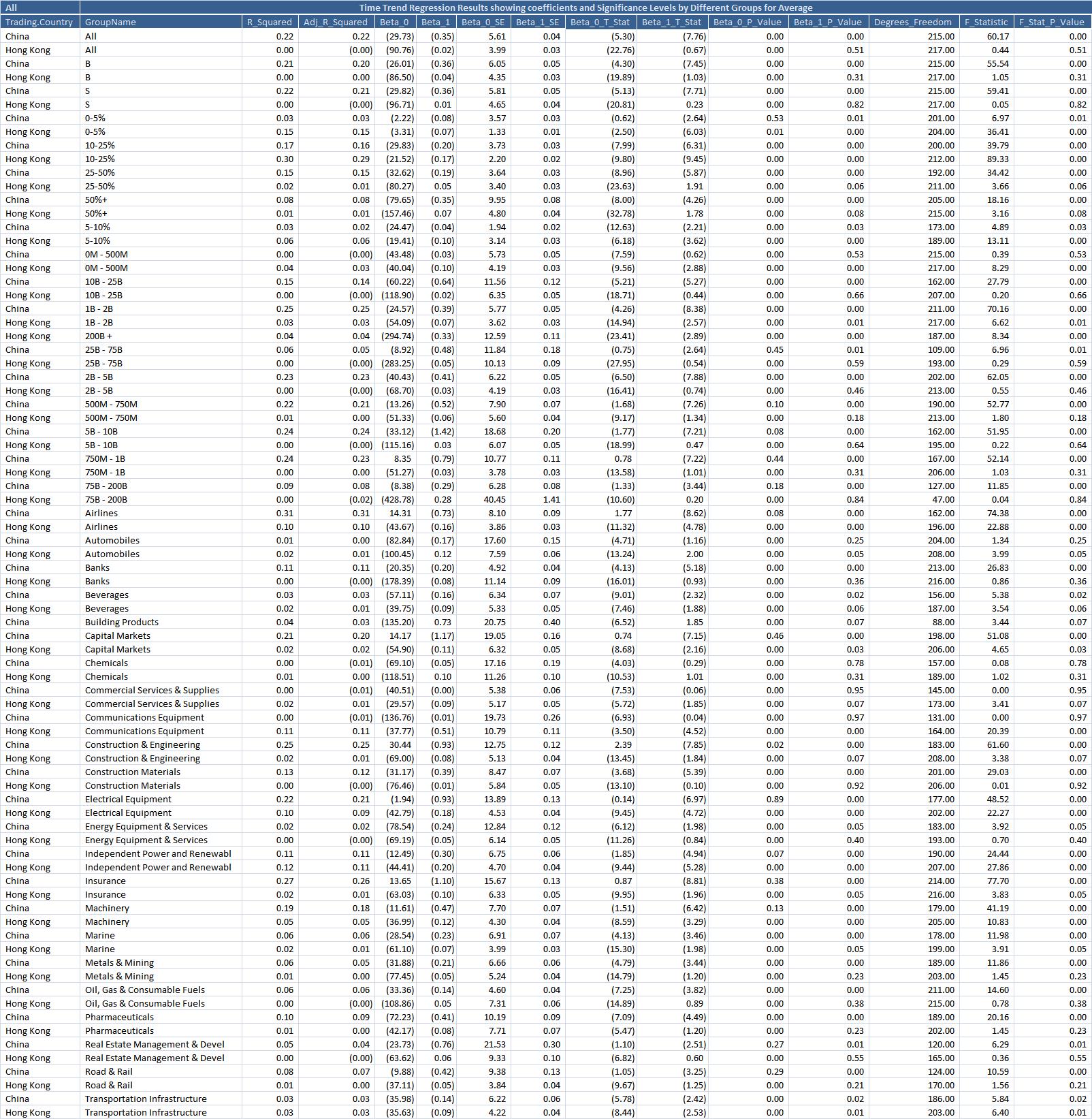}

\caption{Time Trend Regression - Full Sample\label{fig:Time-Trend-Regression-5}}
\end{figure}

\begin{figure}[H]
\includegraphics[width=18cm,height=22cm]{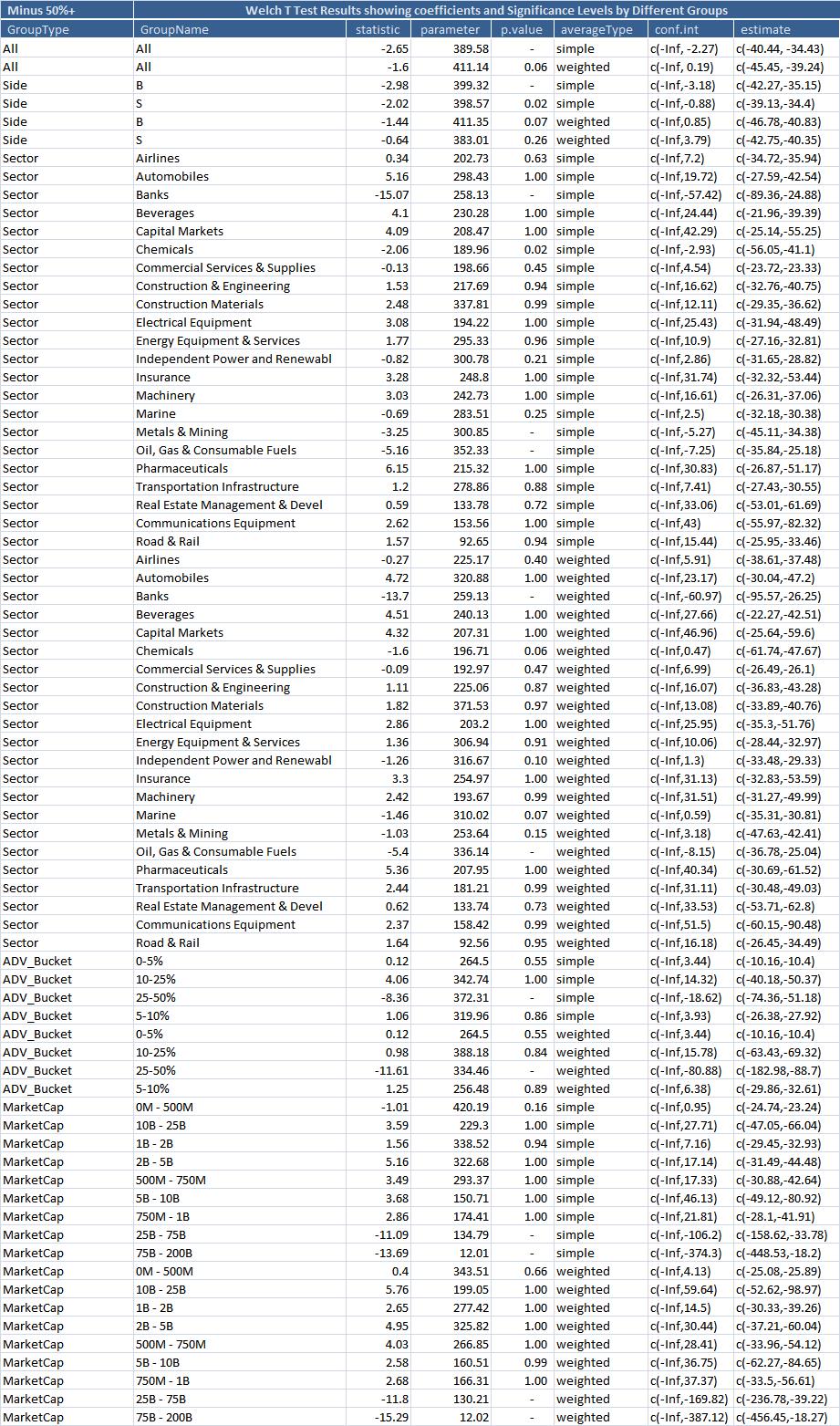}

\caption{Welch T Test - Without 50\%+ ADV\label{fig:Welch-T-Test}}

\end{figure}

\begin{figure}[H]
\includegraphics[width=18cm,height=22cm]{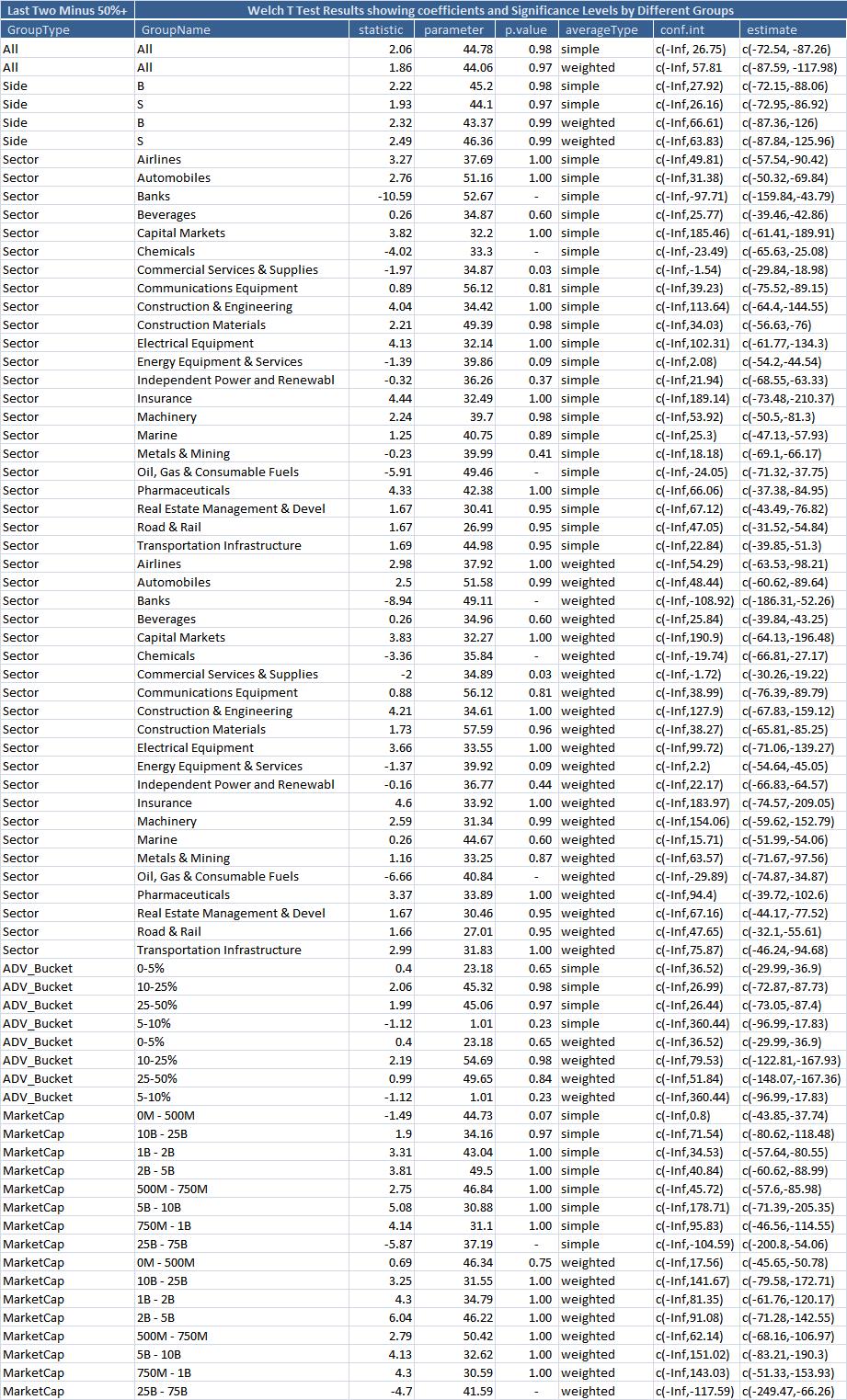}

\caption{Welch T Test - Without 50\%+ ADV Last Two Months\label{fig:Welch-T-Test-1}}
\end{figure}

\begin{figure}[H]
\includegraphics[width=18cm,height=22cm]{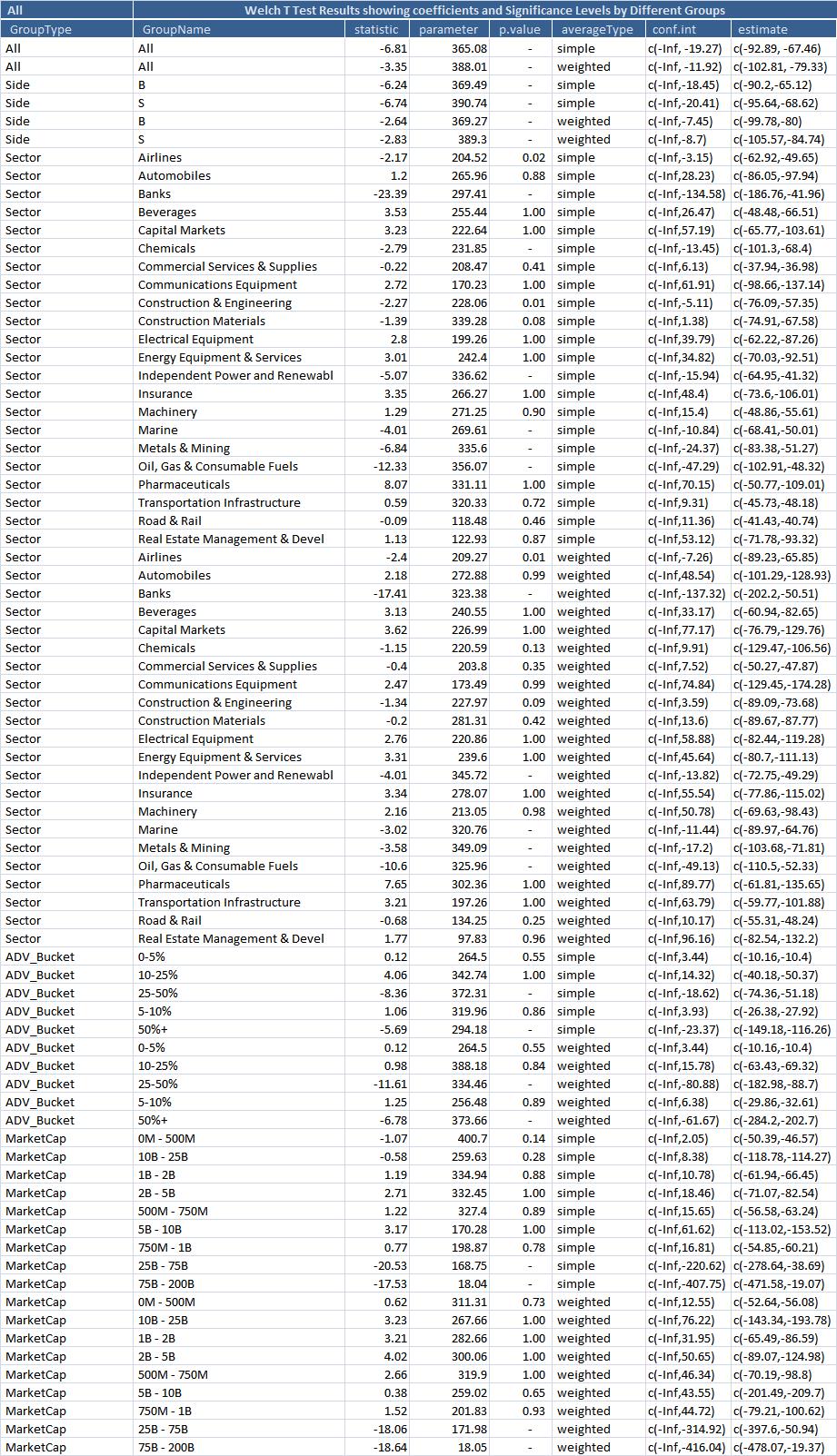}

\caption{Welch T Test - Full Sample\label{fig:Welch-T-Test-2}}
\end{figure}
\begin{figure}[H]
\includegraphics[width=8cm,height=5cm]{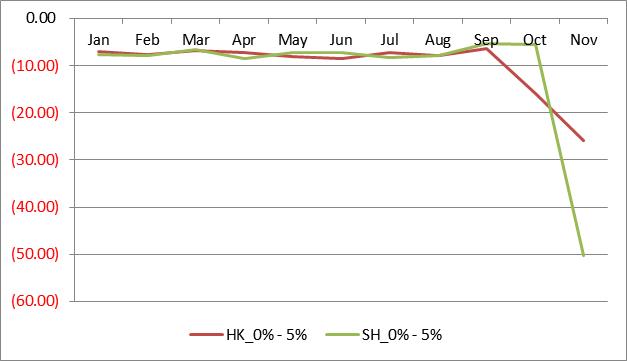}\includegraphics[width=8cm,height=5cm]{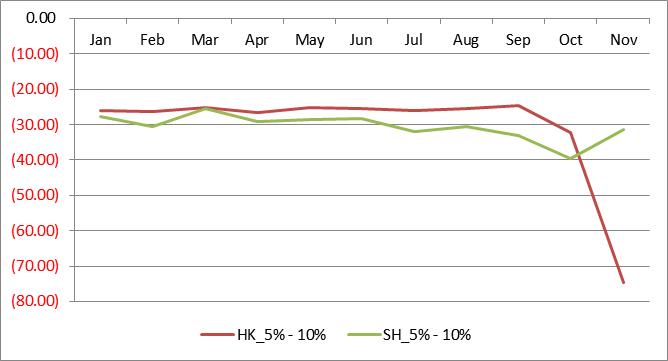}

\includegraphics[width=8cm,height=5cm]{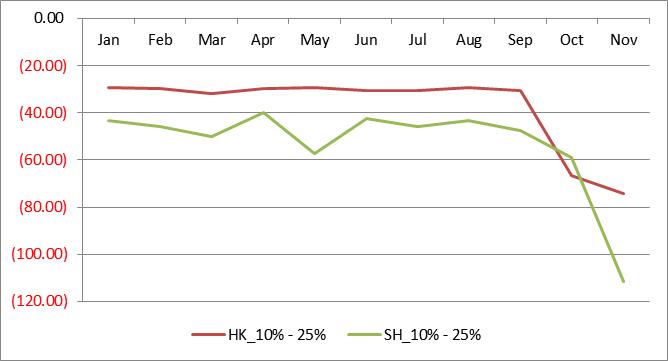}\includegraphics[width=8cm,height=5cm]{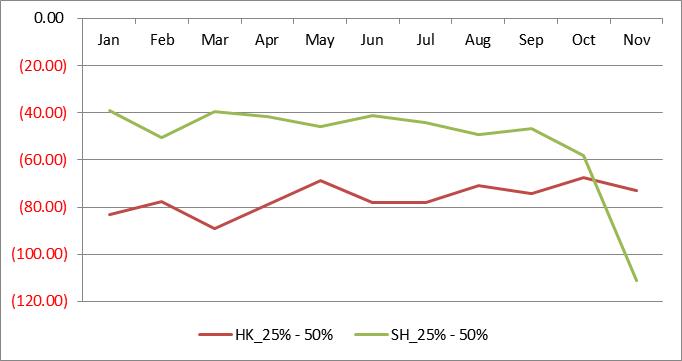}

\includegraphics[width=8cm,height=5cm]{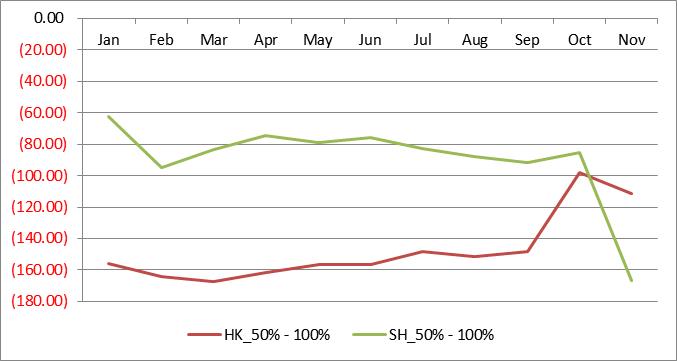}\includegraphics[width=8cm,height=5cm]{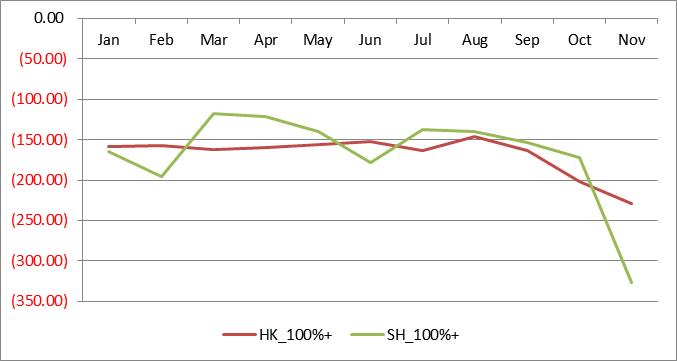}\caption{Simulation Trading Costs by \% ADV Demand\label{fig:Simulation-Trading-Costs-1}}
\end{figure}
The following Market Cap buckets are defined: 
\end{doublespace}
\begin{enumerate}
\begin{doublespace}
\item Small Cap less than 1 Billion USD 
\item Mid Cap 1 Billion to 10 Billion USD 
\item Large Cap 10 Billion USD and above
\end{doublespace}
\end{enumerate}
\begin{doublespace}
\begin{figure}[H]
\includegraphics[width=5.5cm,height=5cm]{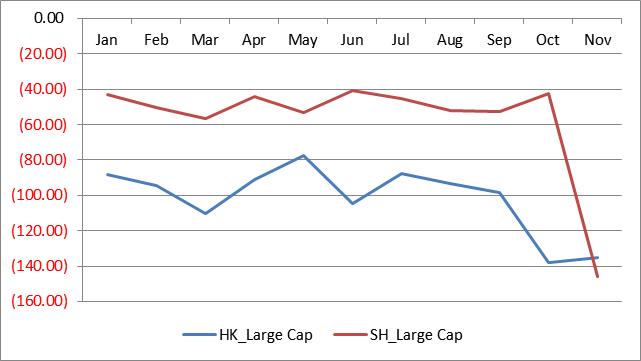}\includegraphics[width=5.5cm,height=5cm]{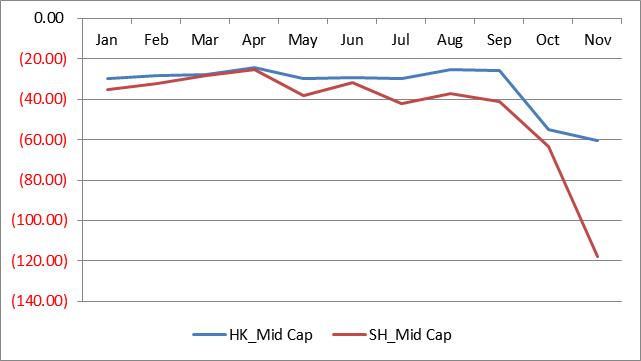}\includegraphics[width=5.5cm,height=5cm]{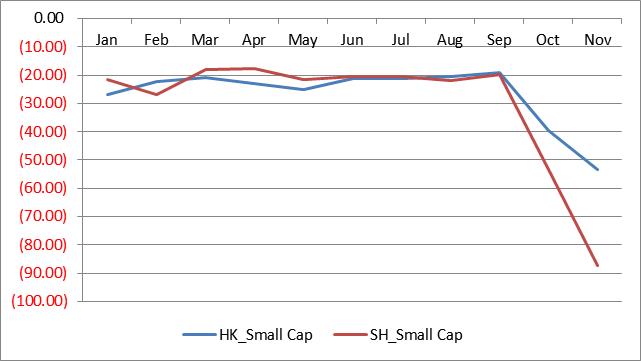}\caption{Simulation Trading Costs by Market Capitalization\label{fig:Simulation-Trading-Costs-2}}
\end{figure}
\begin{figure}[H]
\includegraphics[width=5.5cm,height=5cm]{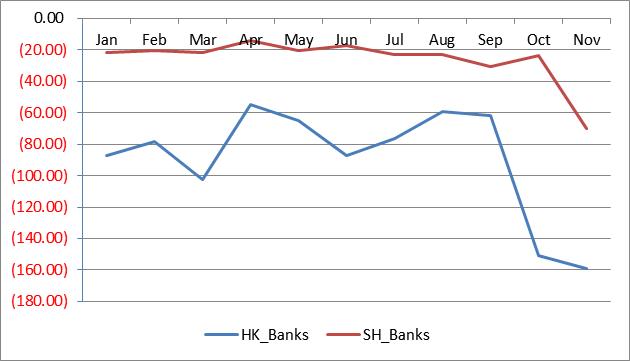}\includegraphics[width=5.5cm,height=5cm]{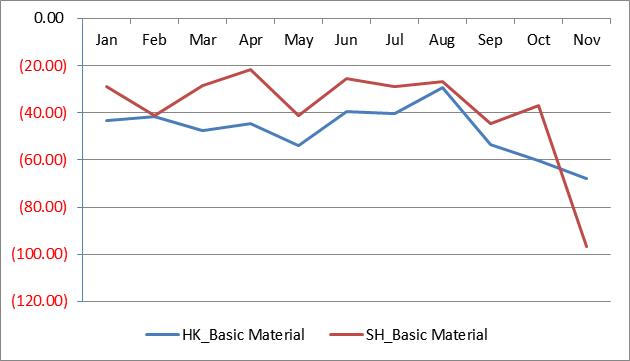}\includegraphics[width=5.5cm,height=5cm]{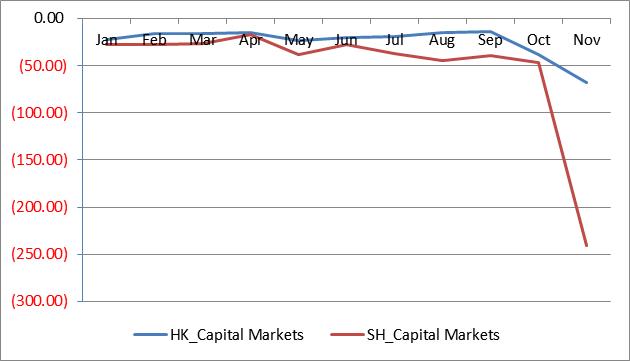}

\includegraphics[width=5.5cm,height=5cm]{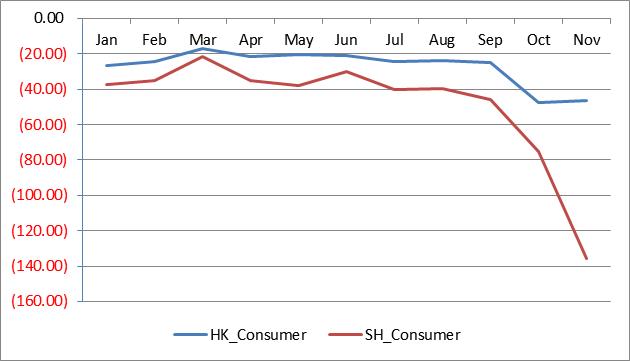}\includegraphics[width=5.5cm,height=5cm]{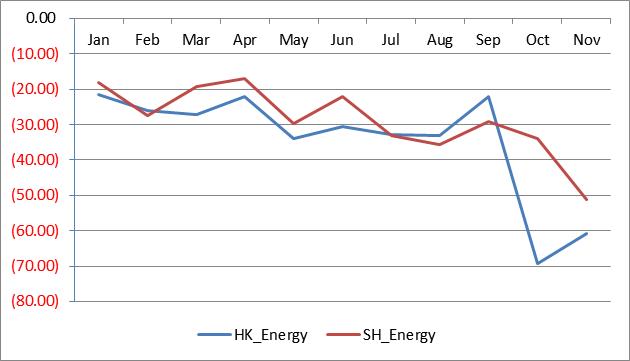}\includegraphics[width=5.5cm,height=5cm]{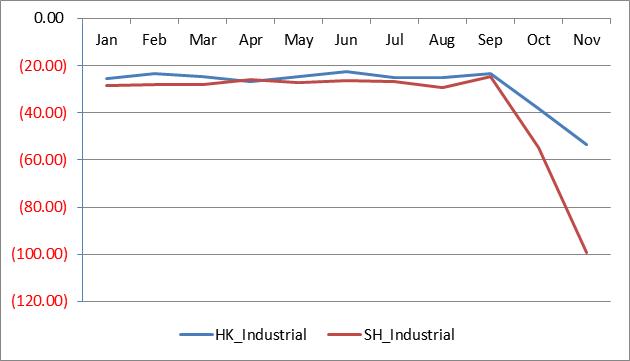}

\includegraphics[width=5.5cm,height=5cm]{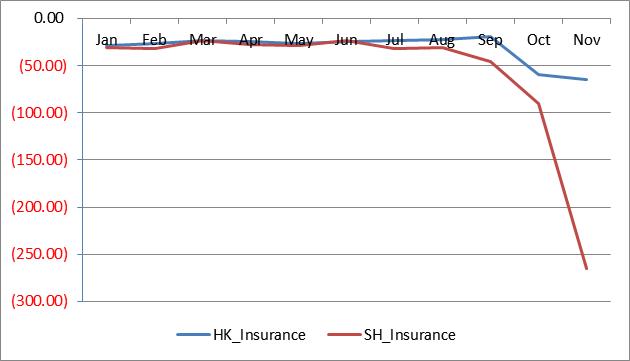}\includegraphics[width=5.5cm,height=5cm]{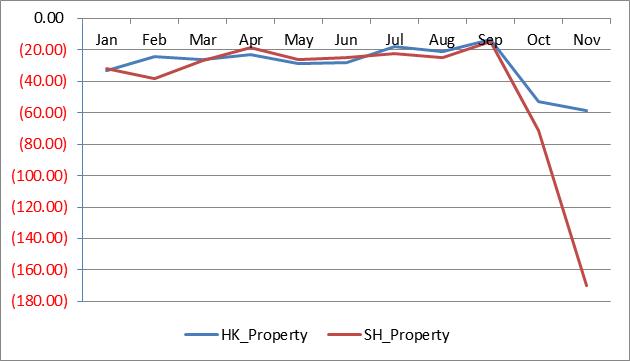}\includegraphics[width=5.5cm,height=5cm]{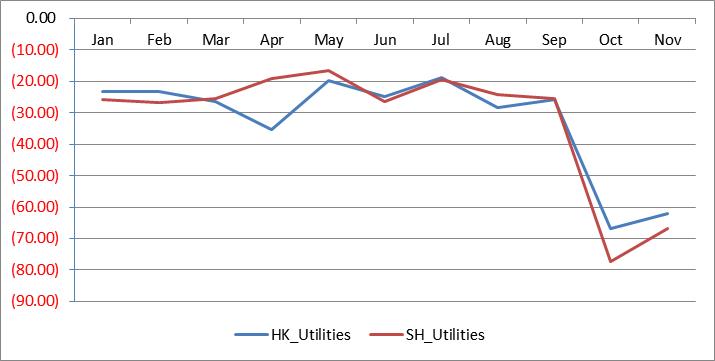}

\caption{Simulation Trading Costs by Sector\label{fig:Simulation-Trading-Costs-3}}
\end{figure}

\end{doublespace}
\begin{doublespace}

\subsection{Comparison of Estimated and Actual Costs on Real Orders\label{subsec:Comparison-of-Estimated}}
\end{doublespace}

\begin{doublespace}
\begin{figure}[H]
\includegraphics[width=18cm]{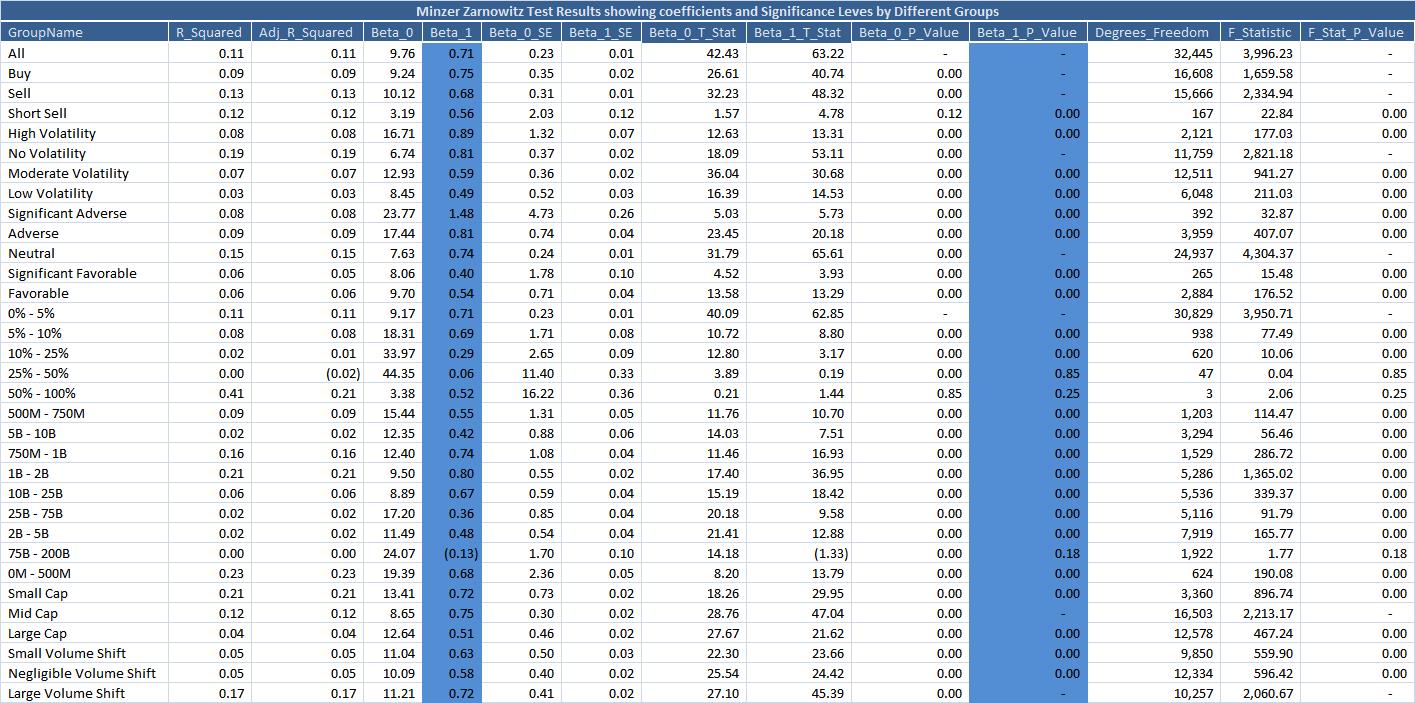}

\caption{Mincer Zarnowitz Regression Results\label{fig:Mincer-Zarnowitz-Regression}}
\end{figure}

\begin{figure}[H]
\includegraphics[width=18cm]{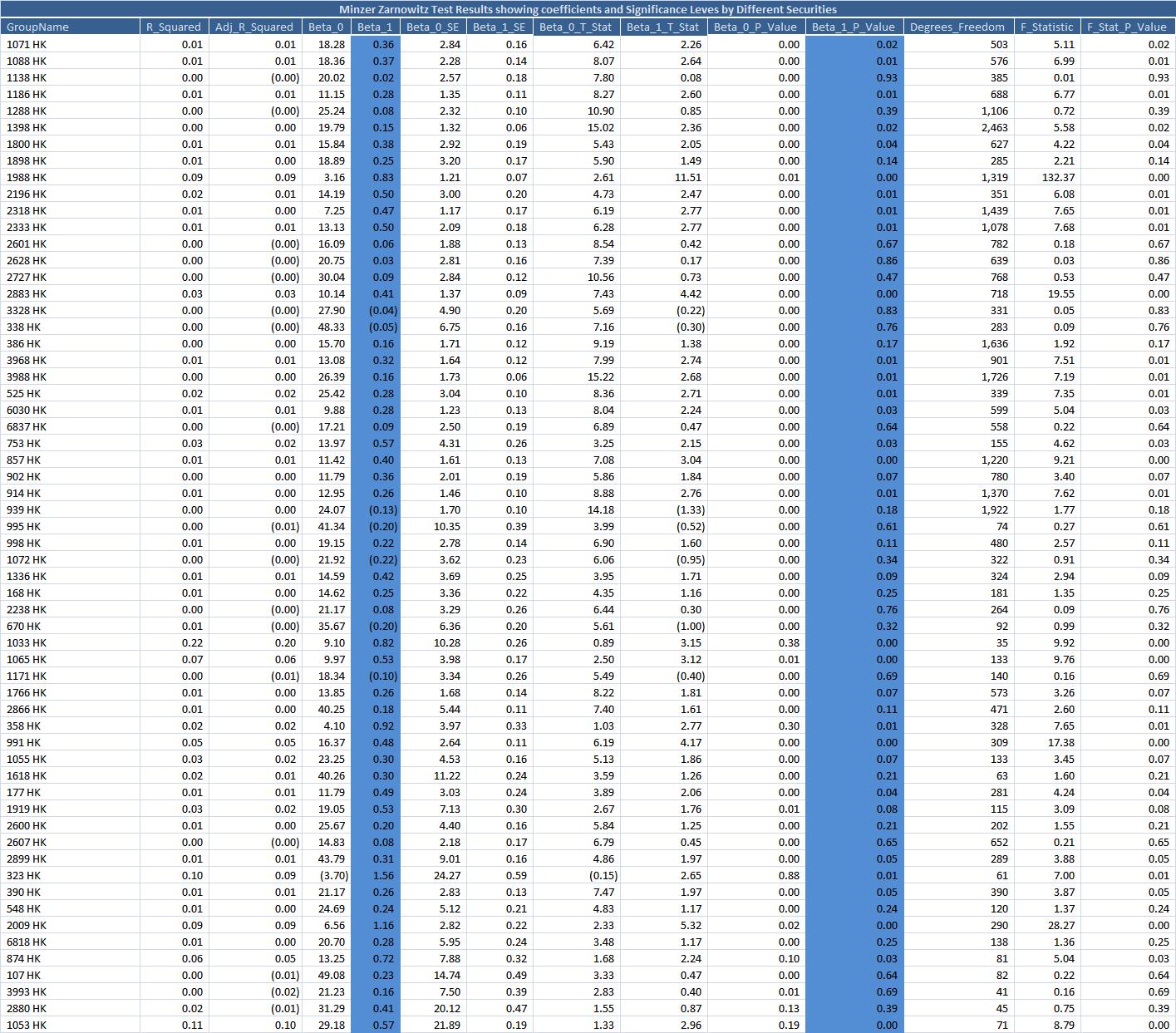}

\caption{Mincer Zarnowitz Regression Results - HK Securities\label{fig:Mincer-Zarnowitz-Regression-1}}
\end{figure}

\begin{figure}[H]
\includegraphics[width=9cm]{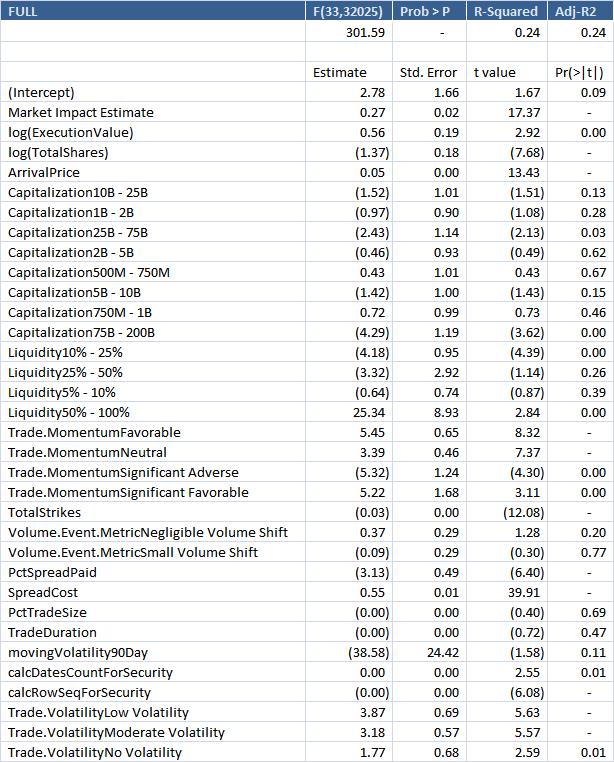}\includegraphics[width=9cm]{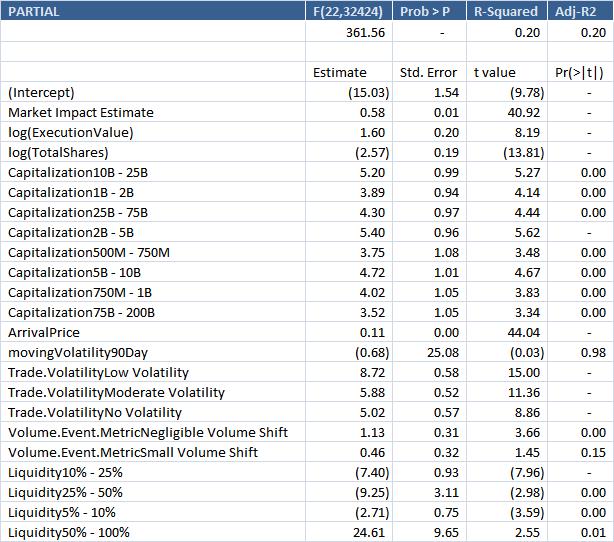}

\caption{Estimate Regression Results \label{fig:Estimate-Regression-Results}}
\end{figure}

\begin{figure}[H]
\includegraphics[width=18cm]{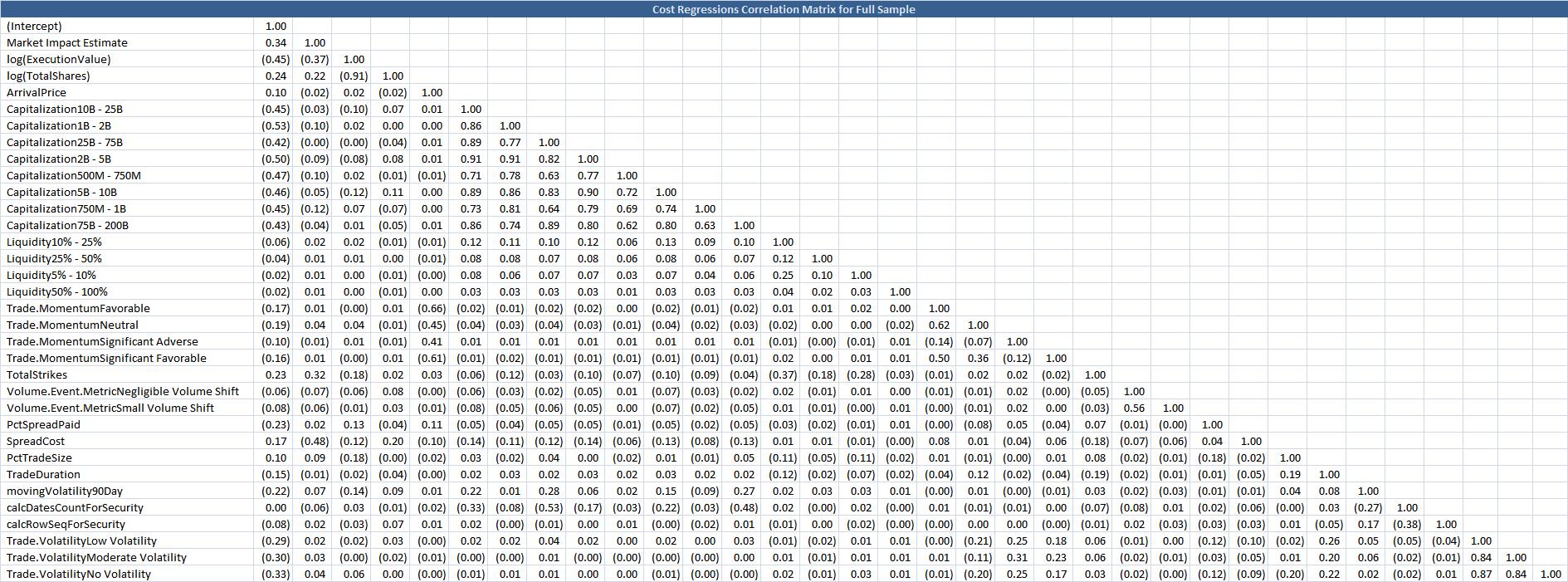}

\caption{Regression Correlation Matrix\label{fig:Regression-Correlation-Matrix}}

\end{figure}

The Five Trade Momentum buckets are based on the side adjusted percentage
return during the order's trading interval:
\end{doublespace}
\begin{enumerate}
\begin{doublespace}
\item Significant Adverse (<-2\%) 
\item Adverse (-1/3\% thru -2\%) 
\item Neutral (-1/3\% thru +1/3\%) 
\item Favorable (+1/3\% thru 2\%) 
\item Significant Favorable (>+2\%) 
\end{doublespace}
\end{enumerate}
\begin{doublespace}
The Four Trade Volatility buckets are based on the coefficient of
variation of prices during the execution horizon: 
\end{doublespace}
\begin{enumerate}
\begin{doublespace}
\item High Volatility (>0.0050) 
\item Moderate Volatility (0.0010 thru 0.0050) 
\item Low Volatility (0.000000000000001 thru 0.0010) 
\item No Volatility (<= 0.000000000000001) 
\end{doublespace}
\end{enumerate}
\begin{doublespace}
The Volume Event Metric (VEM) measure captures the magnitude of the
volume shift on a trading day for a specific stock. We compare a stock’s
current volume profile to the past 60 day’s average profile in each
half-hour interval of the trading day. The absolute values of these
percent of daily volume differences in each interval are then summed
up to create the Volume Event Metric. The below three categorizations
are used: 
\end{doublespace}
\begin{enumerate}
\begin{doublespace}
\item Negligible Volume Shift (VEM < 30\%) 
\item Small Volume Shift (VEM >= 30\% and < 40\%) 
\item Large Volume Shift (VEM >= 40\%) 
\end{doublespace}
\end{enumerate}
\begin{doublespace}
\begin{figure}[H]
\includegraphics[width=5.5cm,height=5cm]{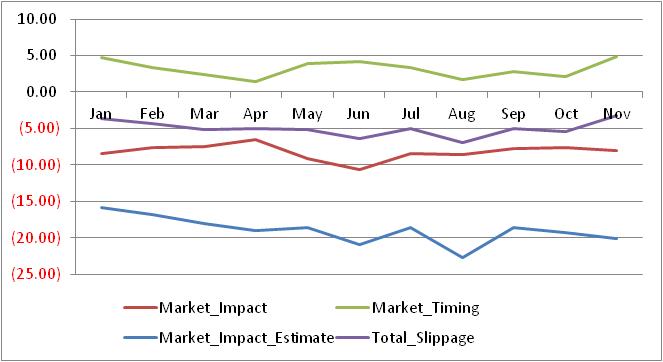}\includegraphics[width=5.5cm,height=5cm]{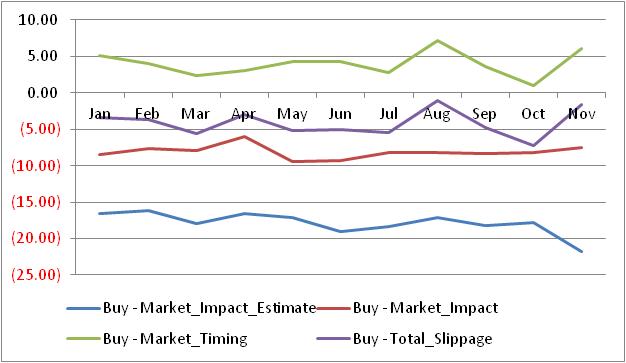}\includegraphics[width=5.5cm,height=5cm]{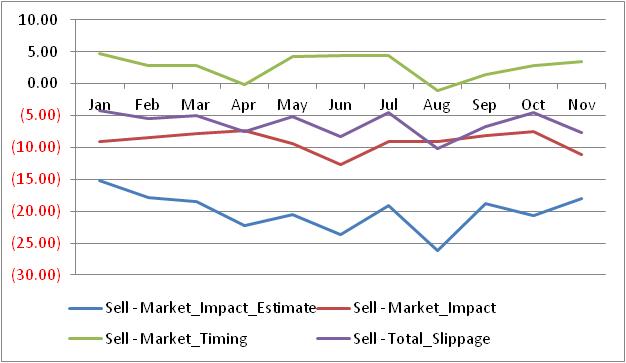}

\caption{Trading Costs on Real Orders\label{fig:Trading-Costs-on}}
\end{figure}
\begin{figure}[H]
\includegraphics[width=8cm,height=5cm]{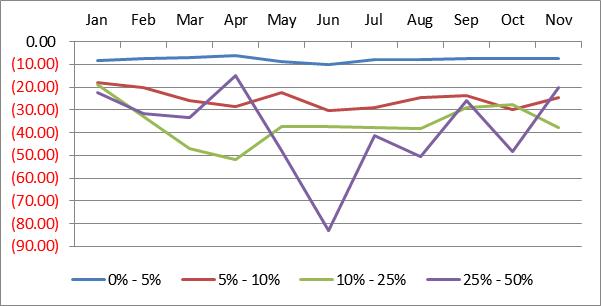}\includegraphics[width=8cm,height=5cm]{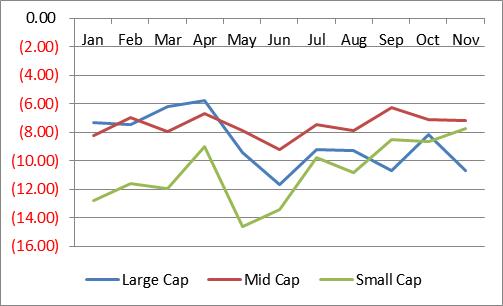}

\includegraphics[width=8cm,height=5cm]{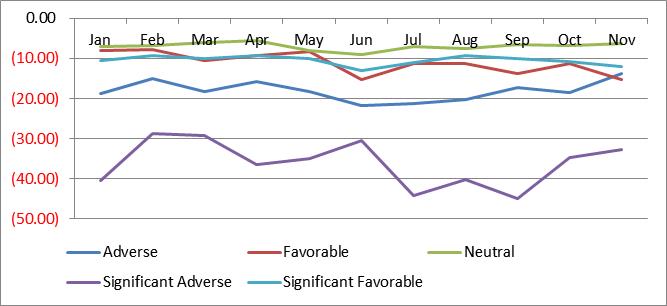}\includegraphics[width=8cm,height=5cm]{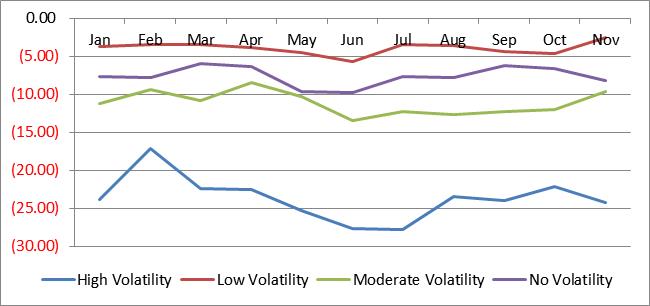}

\includegraphics[width=8cm,height=5cm]{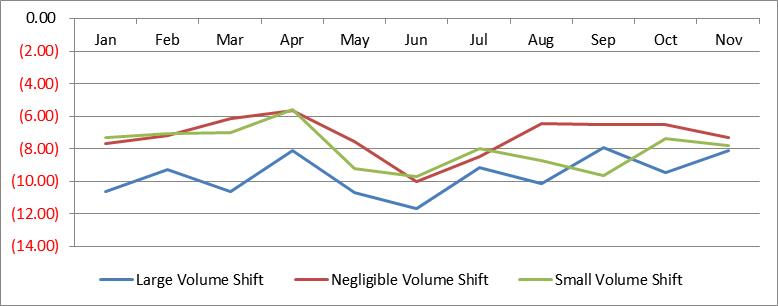}\caption{Market Impact Costs on Real Orders\label{fig:Market-Impact-Costs}}
\end{figure}

\begin{figure}[H]
\includegraphics[width=8cm,height=5cm]{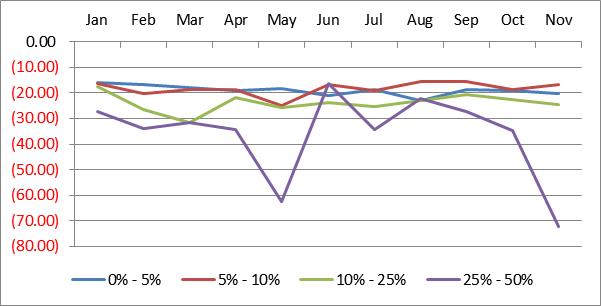}\includegraphics[width=8cm,height=5cm]{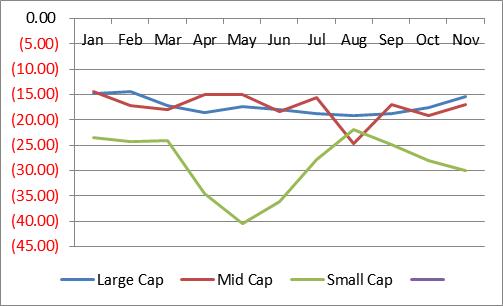}

\includegraphics[width=8cm,height=5cm]{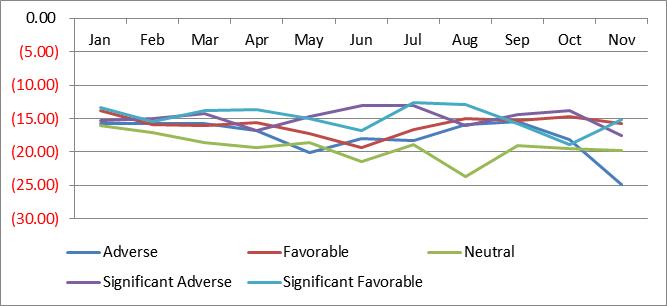}\includegraphics[width=8cm,height=5cm]{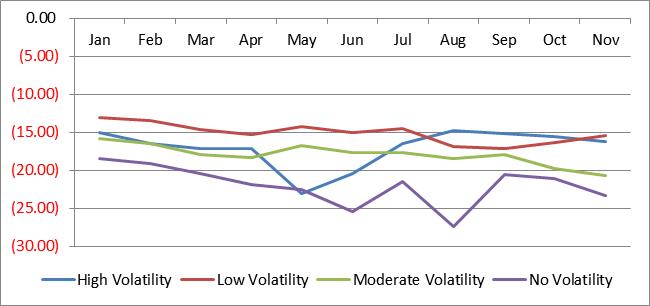}

\includegraphics[width=8cm,height=5cm]{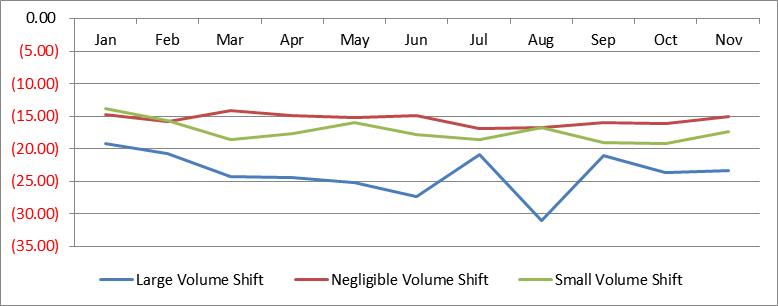}\caption{Market Impact Estimate Costs on Real Orders\label{fig:Market-Impact-Estimate}}
\end{figure}
\begin{figure}[H]
\includegraphics[width=8cm,height=5cm]{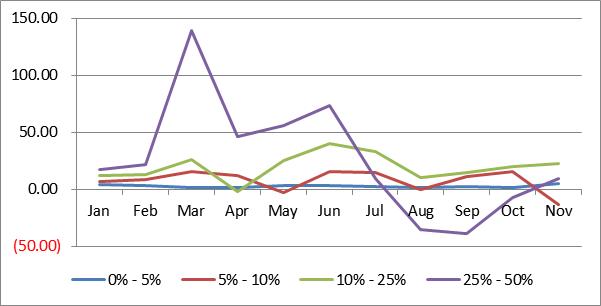}\includegraphics[width=8cm,height=5cm]{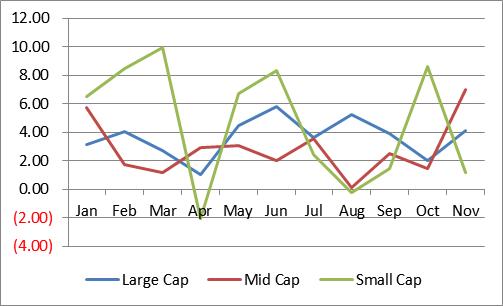}

\includegraphics[width=8cm,height=5cm]{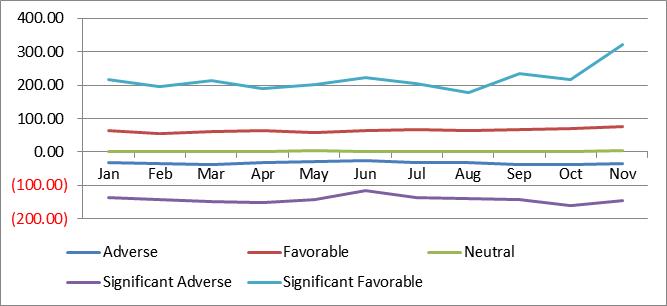}\includegraphics[width=8cm,height=5cm]{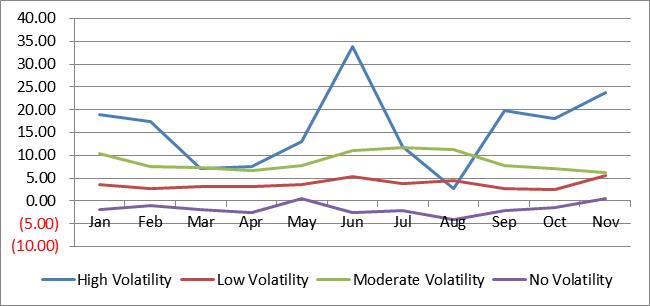}

\includegraphics[width=8cm,height=5cm]{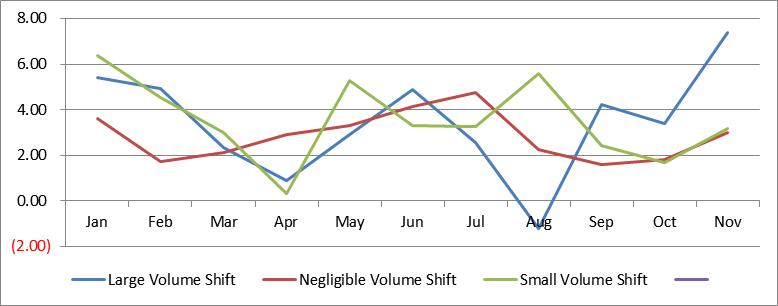}

\caption{Market Timing Costs on Real Orders\label{fig:Market-Timing-Costs}}
\end{figure}
\begin{figure}[H]
\includegraphics[width=8cm,height=5cm]{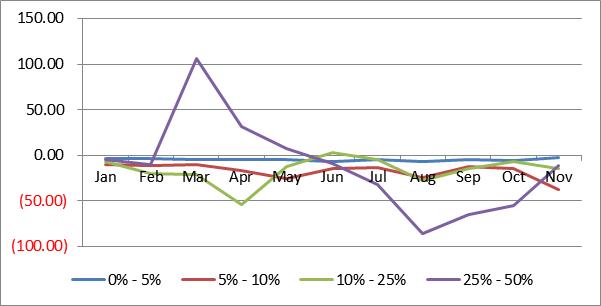}\includegraphics[width=8cm,height=5cm]{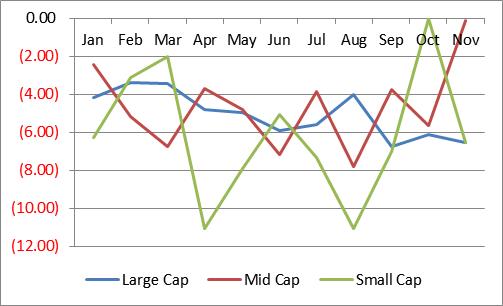}

\includegraphics[width=8cm,height=5cm]{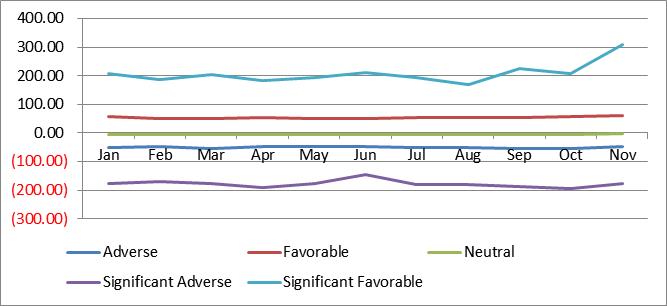}\includegraphics[width=8cm,height=5cm]{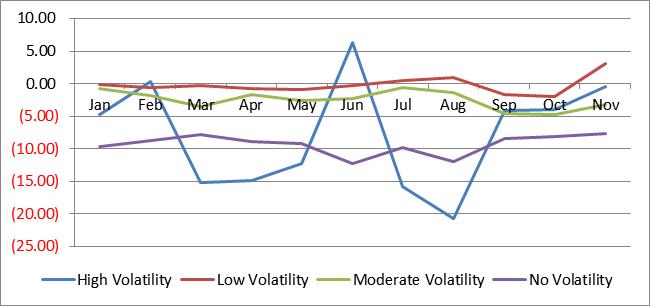}

\includegraphics[width=8cm,height=5cm]{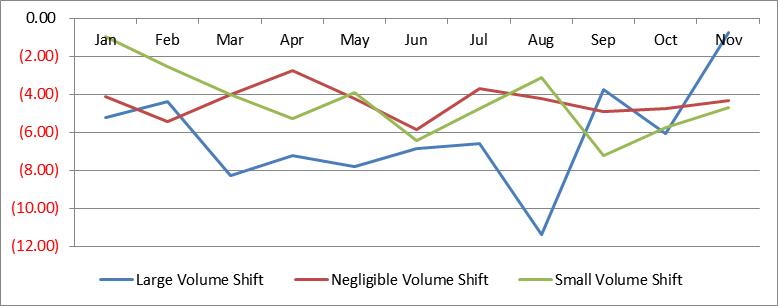}

\caption{Total Slippage Costs on Real Orders\label{fig:Total-Slippage-Costs}}
\end{figure}

\end{doublespace}
\begin{doublespace}

\subsection{Auxiliary Metrics\label{subsec:Auxiliary-Metrics}}
\end{doublespace}

\begin{doublespace}
All the metrics mentioned here are weighted averages, weighted by
the executed value calculated over the same sample of real orders.
Order Duration is in Minutes. The average Trade Size is in number
of shares. Percentage of Spread Paid indicates the percentage of the
spread paid across the order data set. Spread Cost indicates the actual
spread cost in basis points. It is negative here to indicate that
it is a cost.

\begin{figure}[H]
\includegraphics[width=8cm,height=5cm]{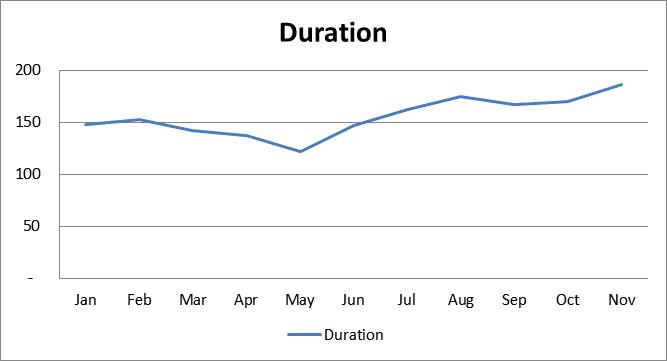}\includegraphics[width=8cm,height=5cm]{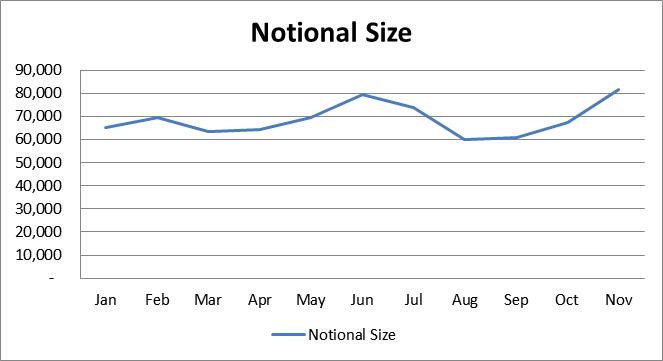}

\includegraphics[width=8cm,height=5cm]{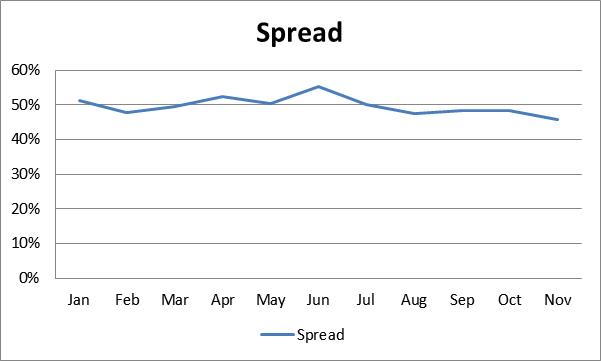}\includegraphics[width=8cm,height=5cm]{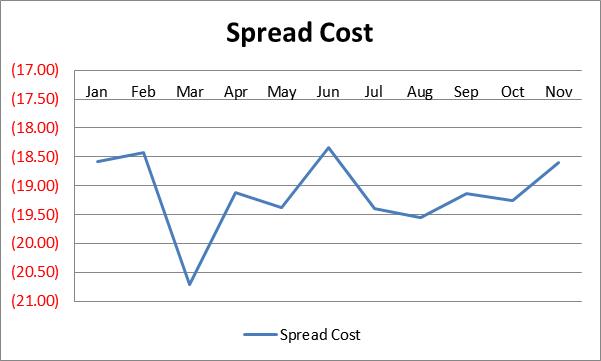}

\includegraphics[width=8cm,height=5cm]{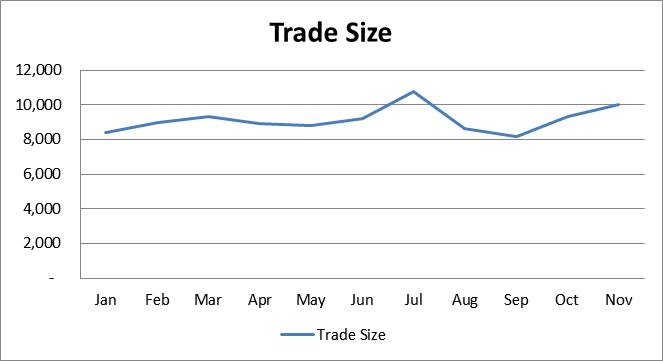}\includegraphics[width=8cm,height=5cm]{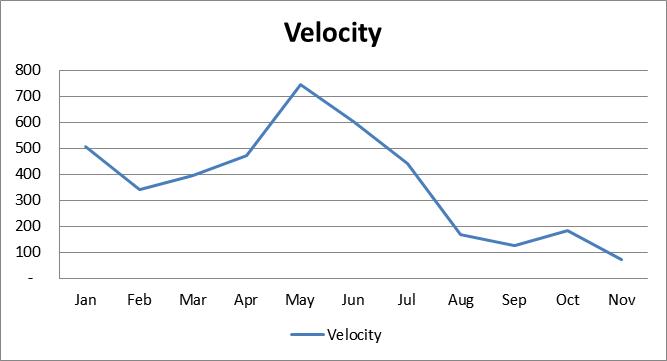}

\includegraphics[width=8cm,height=5cm]{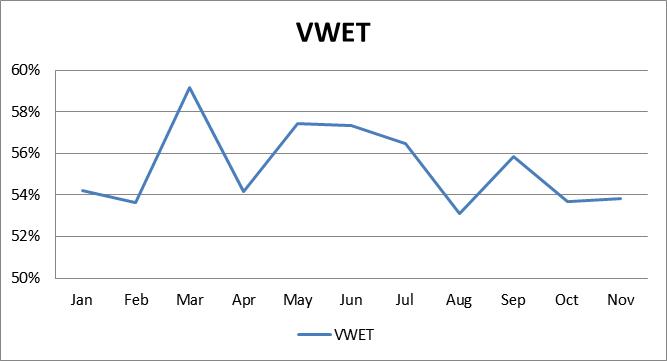}

\caption{Auxiliary Metrics on Real Orders\label{fig:Auxiliary-Metrics-on}}
\end{figure}

\end{doublespace}
\begin{doublespace}

\subsection{Volume Curves\label{subsec:Volume-Curves}}
\end{doublespace}

\begin{doublespace}
\begin{figure}[H]
\includegraphics[width=8cm,height=5cm]{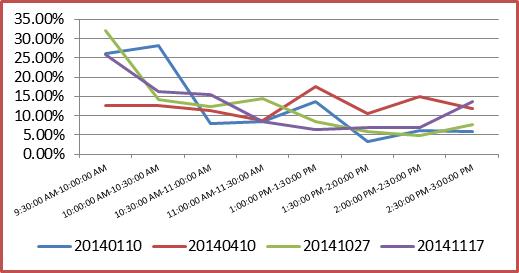}\includegraphics[width=8cm,height=5cm]{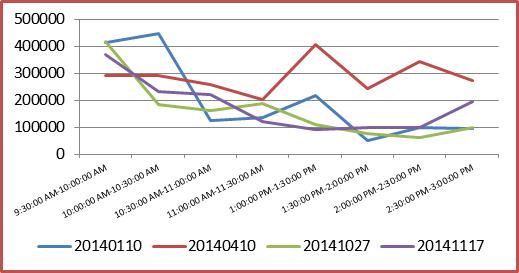}\caption{Volume Curves for 0001.HK\label{fig:Volume-Curves-for}}
\end{figure}
\begin{figure}[H]
\includegraphics[width=8cm,height=5cm]{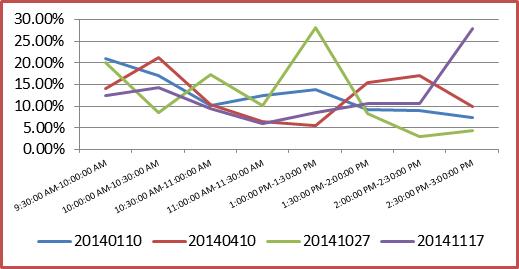}\includegraphics[width=8cm,height=5cm]{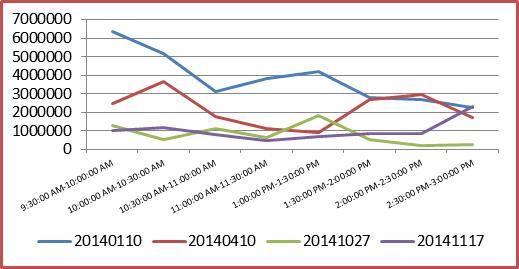}

\caption{Volume Curves for 0688.HK}
\end{figure}

\begin{figure}[H]
\includegraphics[width=8cm,height=5cm]{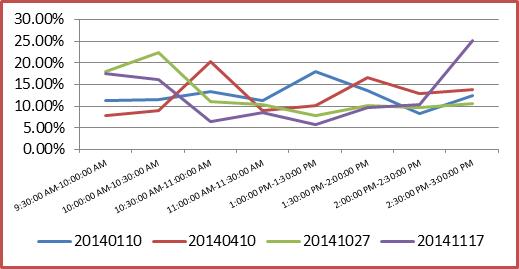}\includegraphics[width=8cm,height=5cm]{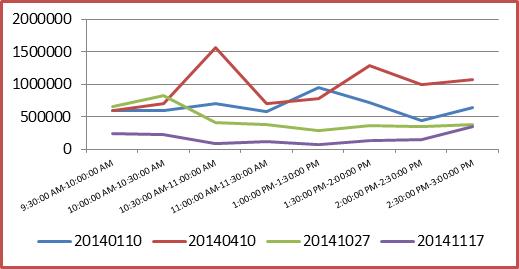}

\caption{Volume Curves for 0813.HK}
\end{figure}

\begin{figure}[H]
\includegraphics[width=8cm,height=5cm]{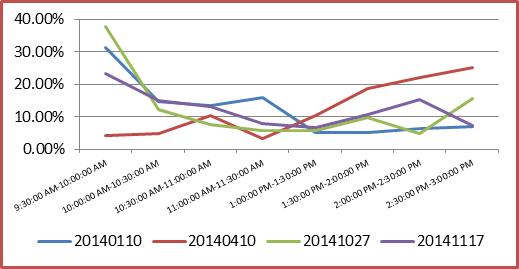}\includegraphics[width=8cm,height=5cm]{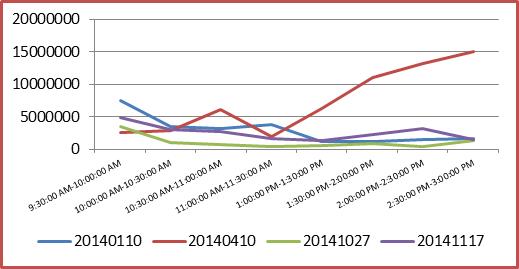}

\caption{Volume Curves for 0992.HK}
\end{figure}

\begin{figure}[H]
\includegraphics[width=8cm,height=5cm]{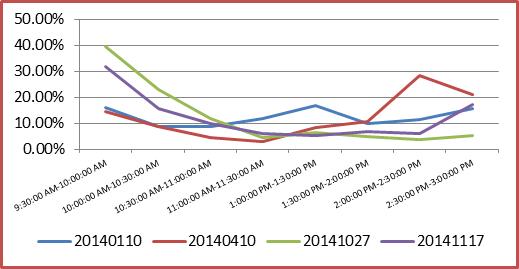}\includegraphics[width=8cm,height=5cm]{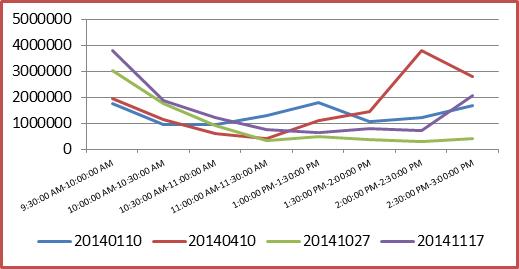}

\caption{Volume Curves for 1928.HK}
\end{figure}

\begin{figure}[H]
\includegraphics[width=8cm,height=5cm]{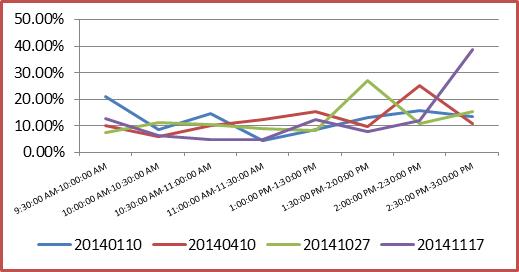}\includegraphics[width=8cm,height=5cm]{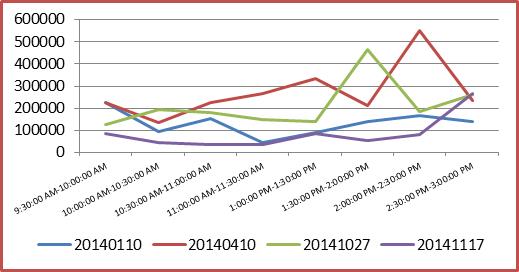}

\caption{Volume Curves for 0669.HK}
\end{figure}

\begin{figure}[H]
\includegraphics[width=8cm,height=5cm]{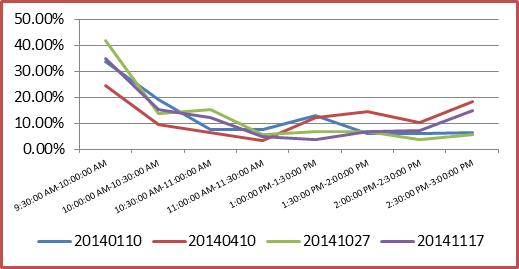}\includegraphics[width=8cm,height=5cm]{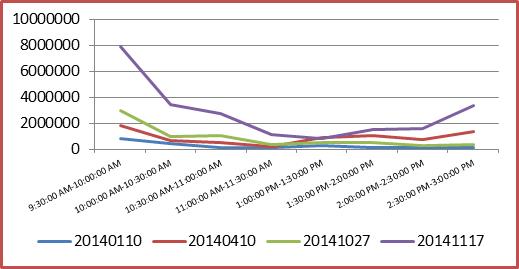}

\caption{Volume Curves for 0700.HK}
\end{figure}

\begin{figure}[H]
\includegraphics[width=8cm,height=5cm]{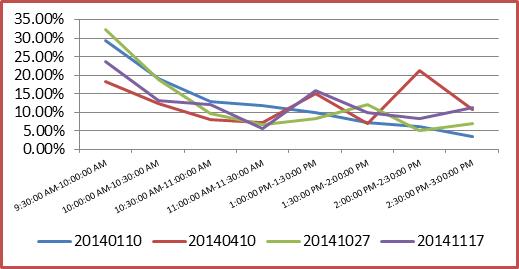}\includegraphics[width=8cm,height=5cm]{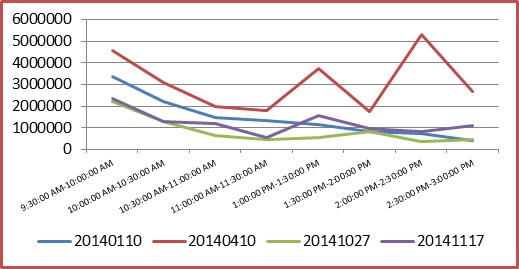}

\caption{Volume Curves for 0941.HK}
\end{figure}

\begin{figure}[H]
\includegraphics[width=8cm,height=5cm]{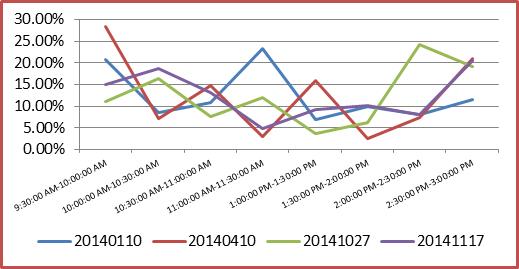}\includegraphics[width=8cm,height=5cm]{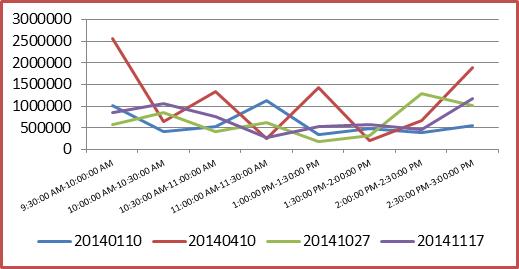}

\caption{Volume Curves for 1114.HK}
\end{figure}

\begin{figure}[H]
\includegraphics[width=8cm,height=5cm]{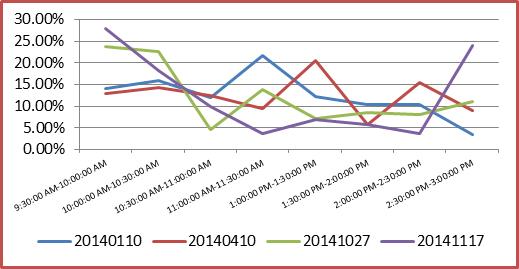}\includegraphics[width=8cm,height=5cm]{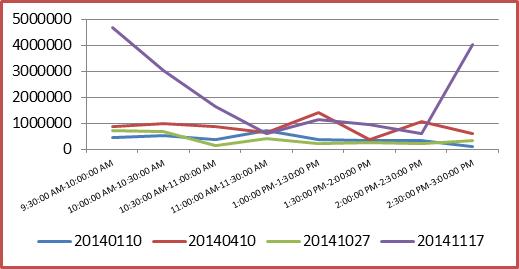}

\caption{Volume Curves for 2388.HK}
\end{figure}

\begin{figure}[H]
\includegraphics[width=8cm,height=5cm]{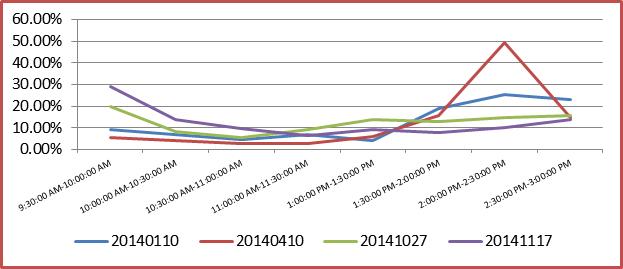}\includegraphics[width=8cm,height=5cm]{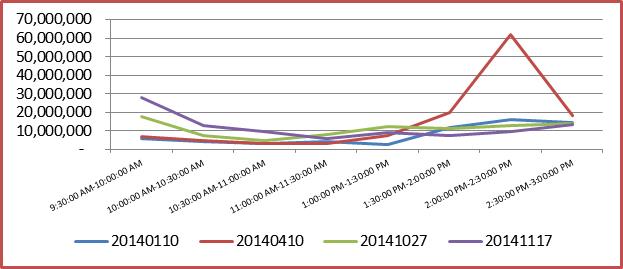}\caption{Volume Curves for 600036.SS}
\end{figure}

\begin{figure}[H]
\includegraphics[width=8cm,height=5cm]{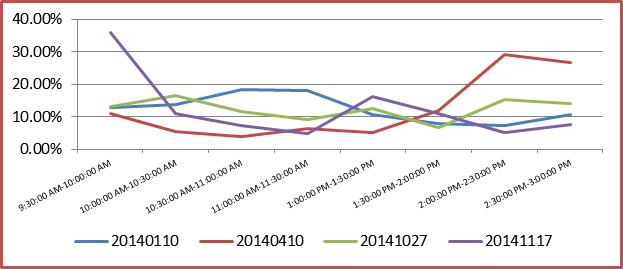}\includegraphics[width=8cm,height=5cm]{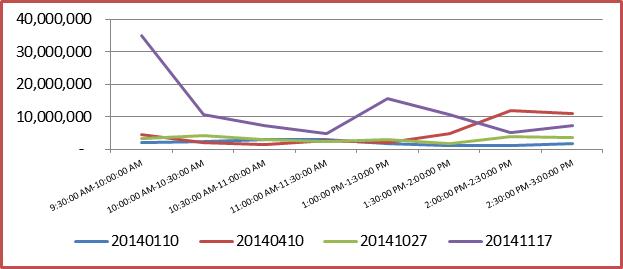}\caption{Volume Curves for 600104.SS}
\end{figure}

\begin{figure}[H]
\includegraphics[width=8cm,height=5cm]{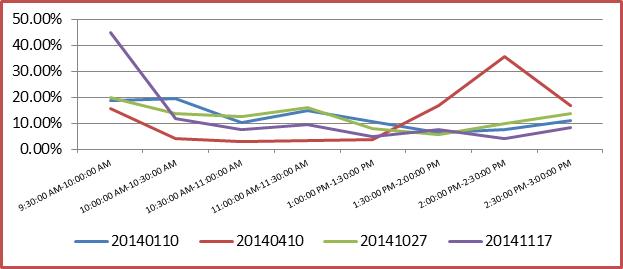}\includegraphics[width=8cm,height=5cm]{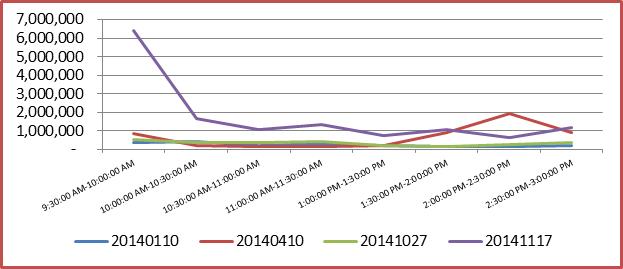}

\caption{Volume Curves for 600519.SS}
\end{figure}

\begin{figure}[H]
\includegraphics[width=8cm,height=5cm]{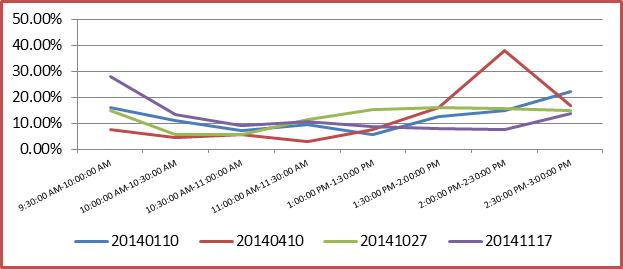}\includegraphics[width=8cm,height=5cm]{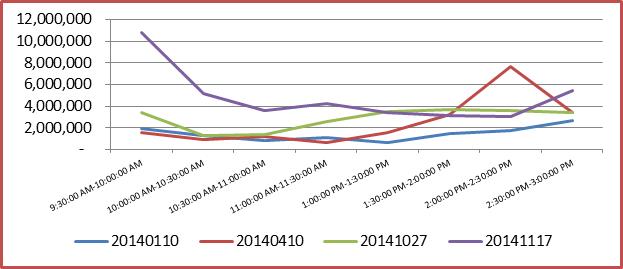}

\caption{Volume Curves for 600900.SS}
\end{figure}

\begin{figure}[H]
\includegraphics[width=8cm,height=5cm]{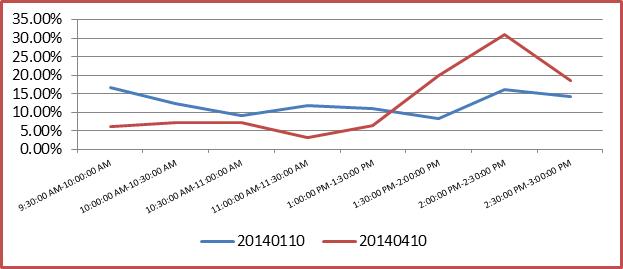}\includegraphics[width=8cm,height=5cm]{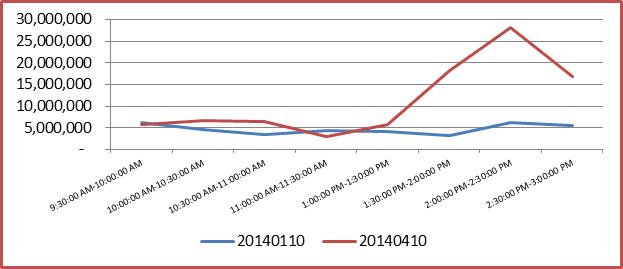}

\caption{Volume Curves for 601299.SS}
\end{figure}

\begin{figure}[H]
\includegraphics[width=8cm,height=5cm]{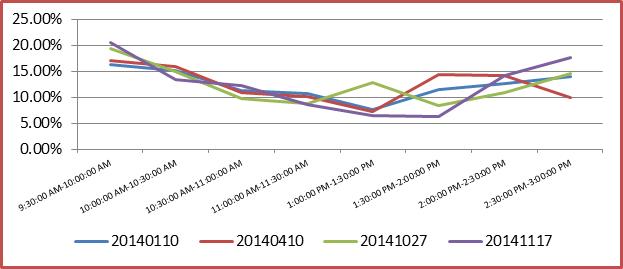}\includegraphics[width=8cm,height=5cm]{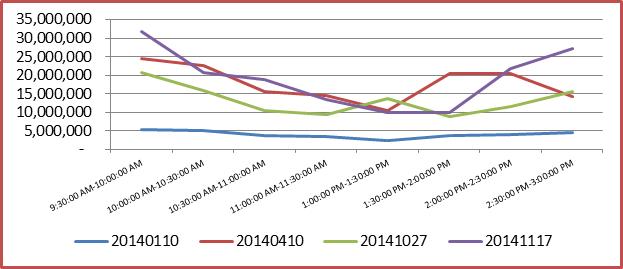}

\caption{Volume Curves for 600048.SS}
\end{figure}

\begin{figure}[H]
\includegraphics[width=8cm,height=5cm]{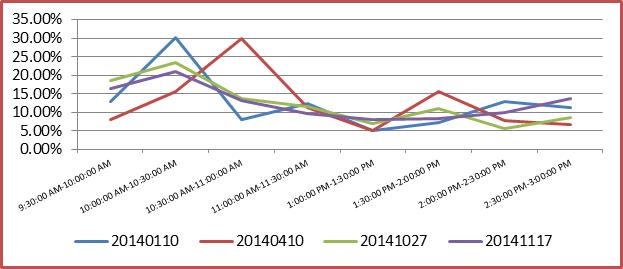}\includegraphics[width=8cm,height=5cm]{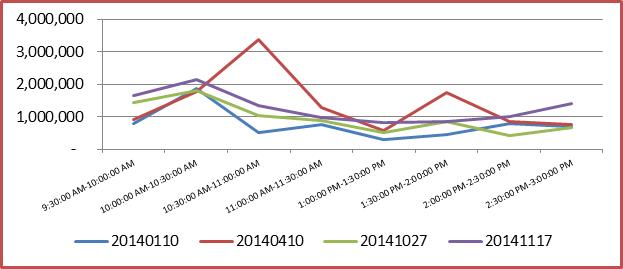}

\caption{Volume Curves for 600372.SS}
\end{figure}

\begin{figure}[H]
\includegraphics[width=8cm,height=5cm]{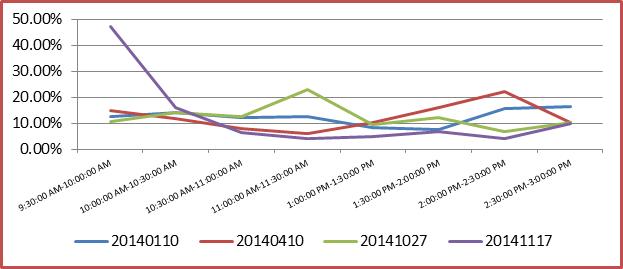}\includegraphics[width=8cm,height=5cm]{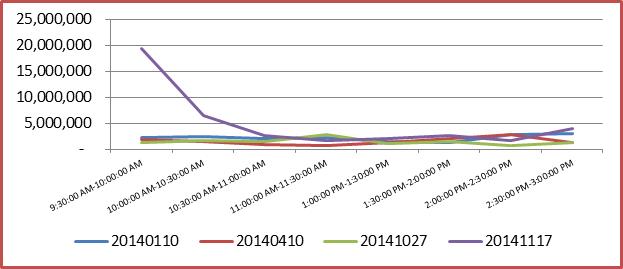}

\caption{Volume Curves for 600690.SS}
\end{figure}

\begin{figure}[H]
\includegraphics[width=8cm,height=5cm]{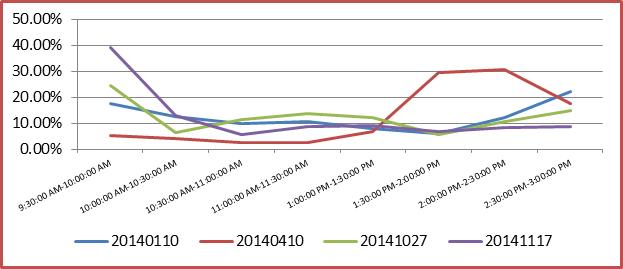}\includegraphics[width=8cm,height=5cm]{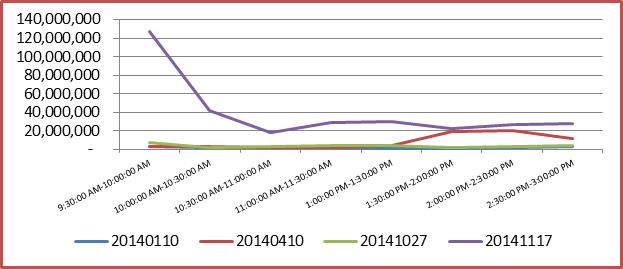}

\caption{Volume Curves for 601006.SS}
\end{figure}

\begin{figure}[H]
\includegraphics[width=8cm,height=5cm]{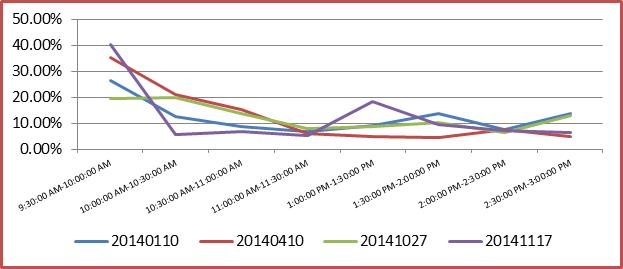}\includegraphics[width=8cm,height=5cm]{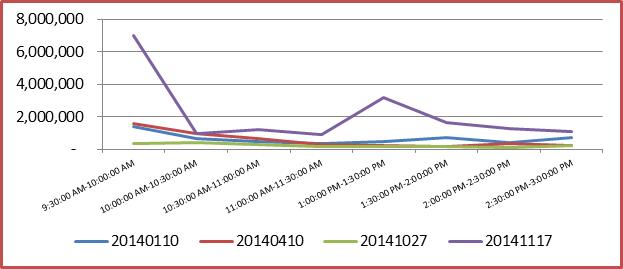}

\caption{Volume Curves for 601888.SS\label{fig:Volume-Curves-for-1}}
\end{figure}
\end{doublespace}

\end{document}